\documentclass[letterpaper]{article} 
\usepackage{algorithm}
\usepackage{algorithmicx}
\usepackage{algpseudocode}
\usepackage{float}
\usepackage{amssymb}
\newtheorem{definition}{Definition}
\usepackage{aaai2026}  
\usepackage{times}  
\usepackage{helvet}  
\usepackage{courier}  
\usepackage[hyphens]{url}  
\usepackage{graphicx} 
\usepackage{natbib}  
\usepackage{caption} 
\usepackage{amsfonts} 
\usepackage{tikz} 
\usetikzlibrary{arrows.meta} 
\usetikzlibrary{positioning} 
\usepackage[table]{xcolor}
\usepackage{booktabs}
\usepackage{amsmath}
\usepackage{multirow}
\usepackage{subcaption}
\usepackage{enumitem}
\usepackage{url}
\usepackage{newfloat}
\usepackage{listings}
\DeclareCaptionStyle{ruled}{labelfont=normalfont,labelsep=colon,strut=off} 
\floatstyle{ruled}
\newfloat{listing}{tb}{lst}{}
\floatname{listing}{Listing}
\title{Who is Helping Whom? Analyzing Inter-dependencies to Evaluate Cooperation in Human-AI Teaming}
\author{
    Upasana Biswas,
    Vardhan Palod,
    Siddhant Bhambri,
    Subbarao Kambhampati
}
\affiliations{
    School of Computing and AI, Arizona State University, Tempe, AZ, USA\\
    $\{\text{ubiswas2,vpalod,sbhambr1,rao}\}$@asu.edu
}
\begin{document}

\maketitle

\begin{abstract}
State-of-the-art methods for Human-AI Teaming and Zero-shot Cooperation focus on task performance, as the sole evaluation metric while being agnostic to `how' the two agents work with each other. Furthermore, subjective user studies only offer limited insight into the quality of cooperation existing within the team. Specifically, we are interested in understanding the cooperative behaviors arising within the team when trained agents are paired with humans - a problem that has been overlooked by the existing literature.
To formally address this problem, we propose the concept of constructive interdependence - measuring how much agents rely on each other’s actions to achieve the shared goal - as a key metric for evaluating cooperation in human-agent teams. We measure interdependence in terms of action interactions in a STRIPS formalism, and define metrics that allow us to assess the degree of reliance between the agents' actions.
We pair state-of-the-art agents with learned human models as well as human participants in a user study for the popular Overcooked domain, and evaluate the task reward and teaming performance for these human-agent teams.
While prior work has claimed that state-of-the-art agents exhibit cooperative behavior based on their high task rewards, our results reveal that these agents often fail to induce cooperation, as evidenced by consistently low interdependence across teams.
Furthermore, our analysis reveals that teaming performance is not necessarily correlated with task reward, highlighting that task reward alone cannot reliably measure cooperation arising in a human-agent team.
\end{abstract}

\begin{links}
    \link{Code}{https://github.com/upasana27/coop-eval-user-study.git}
\end{links}

\section{Introduction}
Achieving zero‑shot cooperation~(ZSC)—i.e., enabling agents to collaborate effectively with previously unseen partners—is a key challenge in cooperative AI. This capability is particularly important in human-agent teaming~(HAT) where agents must interact with and adapt to a diverse range of human behaviors.
Popular approaches for developing such agents often rely on the task reward as the primary signal for learning and evaluation~\citep{metric1,metric2,hsp,fcp}. 
However, task reward alone is often insufficient to fully capture collaborative behavior, as it can obscure important details about individual teammate performance and the interactions that arise between them. 
Agents may learn to optimize task success by operating independently, without adapting to their teammate’s actions. 
This issue is especially pronounced in domains where cooperation is not strictly required for task completion,  as strong task performance may be attained with minimal or no coordination.
\begin{figure}[]
  \centering
        \includegraphics[width=0.7\columnwidth]{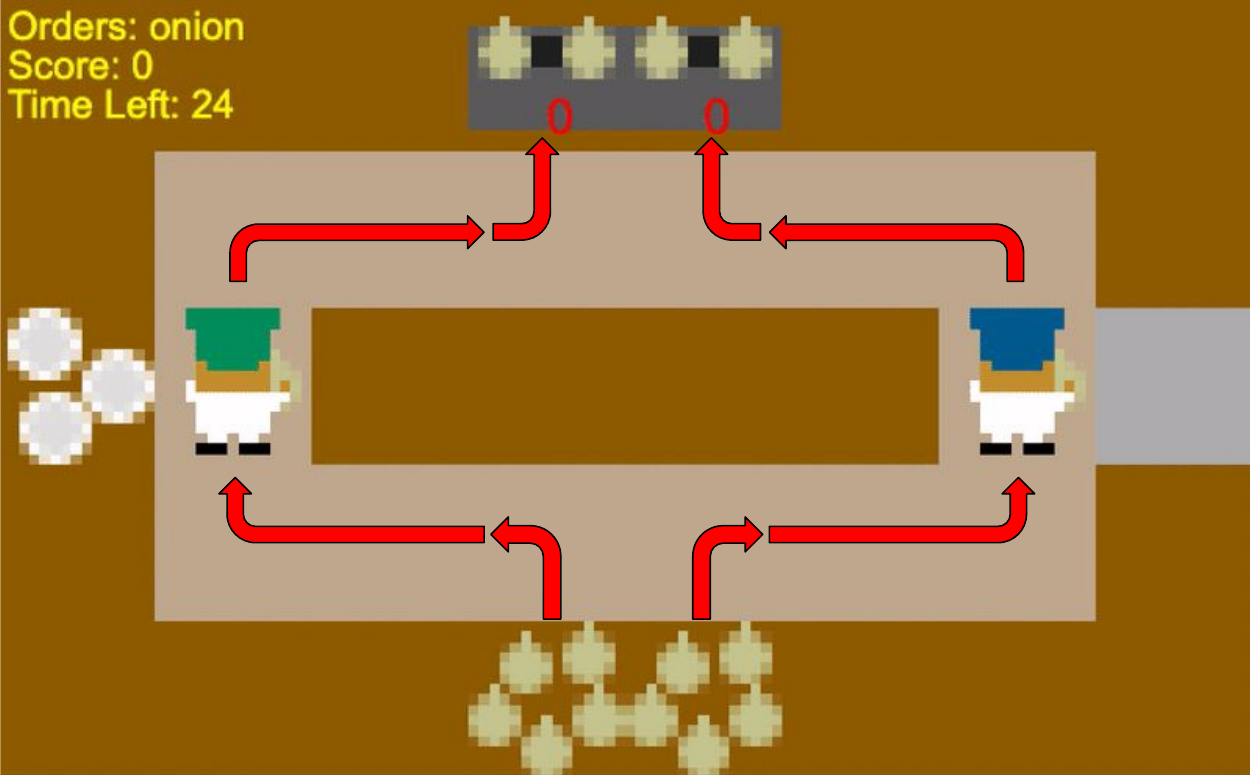}
        \includegraphics[width=0.7\columnwidth]{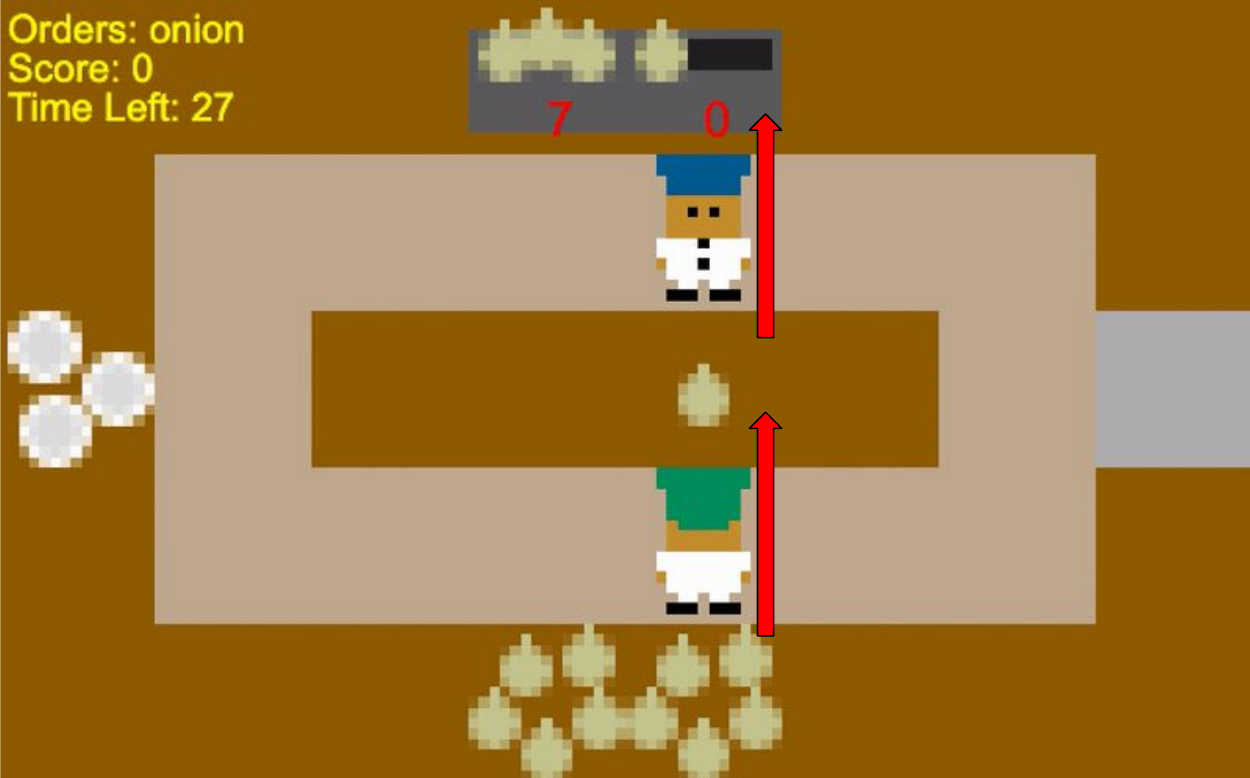}
         \caption{Depicted are two strategies to fill a pot with onions in a cooking game. The coordinated strategy (right) is more efficient than the individual strategies (left), but runs the risk of failure if cooperation is not achieved. }
         \label{fig:coor-example}
\end{figure}

To illustrate this, we draw on the distinction between Required Cooperation~(RC) and Non-Required Cooperation~(Non-RC) settings introduced by~\citet{rc}. In RC scenarios, all team members must participate to achieve the shared goal, whereas in Non-RC settings, individuals can independently complete the task without relying on teammates. For example, consider a human-agent team playing Overcooked (Figure~\ref{fig:coor-example}), where players must prepare and serve soups by collecting and cooking onions.
In this environment, collaboration is not strictly necessary—one player can complete the task while the other remains idle, or the two can engage in active coordination, such as passing onions. Crucially, both strategies can yield similar task rewards despite vastly different levels of cooperation and adaptation. This issue is especially concerning in human-agent teams involving ZSC agents, where humans may compensate for an idle or non-adaptive agent, resulting in high team task rewards that conceal deficiencies in the agent’s cooperative abilities.

The insufficiency of task reward to evaluate ZSC agents is rooted in the shadowed equilibrium problem~\citep{matignon,fulda}, which arises in Non-RC domains with multiple viable strategies to accomplish a task. 
In such settings, agents frequently converge to non-cooperative equilibria because cooperative strategies are rarely explored during training~\citep{obs}. 
As a result, ZSC agents may fail to recognize and reciprocate cooperative intentions when paired with partners who actively seek collaboration~\citep{carroll}.
Evaluating agents solely through task reward conceals these failures of ZSC agents, which are especially detrimental in human-agent teams where unreciprocated cooperation can degrade trust, team cohesion, and overall effectiveness~\citep{trust1,trust2}.
Consequently, relying solely on task reward limits the development and assessment of agents’ ability to adapt and effectively cooperate with diverse human teammates in real-world settings.

To assess cooperation arising in human-agent teams, we focus on \textbf{interdependence}—a structured form of cooperation where team members' actions are contingent on one another~\citep{int1}. This type of coordination is central to many real-world HAT domains such as Urban Search and Rescue~\citep{usar}, collaborative trash collection~\citep{trash}, and predator-prey systems~\citep{pred-prey1, pred-prey2}. While prior work categorizes interdependence into pooled, sequential, reciprocal, and team-based~\citep{Verhagenetal2022,singh1,singh2}, we specifically target sequential and reciprocal interdependencies. We propose a novel metric for measuring such interdependencies between a human and an agent working as a team. We map a two-player Markov game to a symbolic STRIPS-like formalism~\citep{strips}, introducing symbolic structure to both world states and actions.

We pair state-of-the-art ZSC agents with human partners and models~\citep{zsceval}.
We evaluate these teams using the proposed metric in the Overcooked domain—a popular benchmark where numerous approaches have been developed for zero-shot cooperation and HAT~\citep{fcp,carroll,mep,hsp,cole}, thus making it an ideal testbed. The proposed metric generalizes to any domain that can be represented using a symbolic formalism. 
To support ease of integration of the proposed metric into other domains, we provide a fully modular software package (pseudocode in Technical Appendix A), and present a concrete application to a Search and Rescue domain (Technical Appendix B). Our main contributions are:
\begin{itemize}[nosep]
\item We propose a domain-agnostic metric that quantitatively captures sequential and reciprocal interdependence in human-agent teams.
\item Using our proposed metric in the Overcooked domain, we demonstrate that task reward alone can be misleading as an indicator of cooperation in human-agent teams.
\item We validate our metric with a user study and simulations involving state-of-the-art ZSC agents, revealing their limitations in adapting to cooperative partners.
\end{itemize}
Our analysis reveals a critical gap in current ZSC agents: when paired with partners actively seeking collaboration, these agents often fail to adapt. This issue is particularly pronounced in Non-RC settings, where teams may still achieve high task rewards despite minimal genuine coordination. These findings highlight the limitations of relying solely on task reward and emphasize the importance of explicit interdependence metrics to accurately assess cooperation in human-agent teams.
We also present preliminary experiments showing that incorporating the proposed cooperation metric during training leads SOTA ZSC agents to consistently increase constructive interdependencies without harming task performance, suggesting that the metric can effectively promote cooperative behavior.
\section{Related Works}
Prior work in human-agent teaming often relies on task performance or episodic reward as evaluation metrics~\citep{fcp,hsp,mep,cole,zsceval,metric1}. Other approaches emphasize alternative metrics such as collaborative fluency, robot and human idle time~\citep{metric3,metric6,metric7}, or use subjective user studies to assess trust, engagement, and fluency~\citep{metric3,metric4,metric5}. However, these metrics are typically sensitive to specific environment layouts and task structures. Moreover, subjective evaluations offer limited insight into the underlying quality of cooperation. In contrast, our proposed interdependence metric provides a domain-general and objective measure of cooperative behavior between teammates.

\citet{zhang2024} evaluate team outcomes and collaboration characteristics such as contribution rate, individual effort, and communication frequency. \citet{bishop} propose action-based metrics like Productive Chef Actions (PCA), PCA duration, and Chef Role Contribution (CRC) to quantify individual roles during execution. Similarly, \citet{ries2024} compute team contribution by comparing the proportion of tasks completed by humans and AI agents. While these approaches quantify task participation, they do not capture the underlying cooperative dynamics or structural dependencies between teammates’ actions.

Our interdependence metric complements recent work by \citet{wang2025}, which focuses on how human-agent teams adapt and evolve over time. While shared goals and team acceptance are often measured via subjective reports~\citep{liang2019}, our metric offers an objective signal of cooperative behavior by revealing when teammates act in ways that enable or anticipate each other’s contributions. Successfully formed and fulfilled interdependencies can indicate role adherence~\citep{wang2024}, trust, and coordination~\citep{moran2023,cai2019}.

Interdependence has also been positioned as a central organizing principle for human-machine teaming~\citep{int5,int1}. Prior work has explored coordination through low-level dynamics using Convergent Cross Mapping (CCM)~\citep{barton2018,pred-prey1,pred-prey2}, whereas our approach targets symbolic, structured task dependencies. \citet{Verhagenetal2022,singh2} categorize interdependence into pooled, sequential, reciprocal, and team-based. We formalize interdependence when one agent’s action effect satisfies another’s precondition, using a STRIPS-based representation. This captures both unidirectional (sequential) and bidirectional (reciprocal) dependencies in human-agent collaboration.
\section{Preliminaries}
\subsection{Two-Player Markov Game}
A two-player Markov game for a human-AI cooperation scenario can be defined as $\langle S, A, T, R \rangle$ where $S$ is the set of world states, $A : A_{1} \times A_{2}$ where $A_{i}$ is the set of possible actions for agent $i$, $T : S \times A_{1} \times A_{2} \rightarrow S$ is the transition function mapping the present state and the joint action of the agents to the next state of the world, and $R_{i} : S \times A_{1} \times A_{2} \rightarrow R_{i}$ is the reward function mapping the state of the world and the joint action to the global reward.

For a 2-player cooperative Markov game, $R = R_{1} = R_{2}$ where $R$ is the global environment reward function. The joint policy is defined as $\pi = \left( \pi_{1}, \pi_{2} \right)$ where the policy $\pi_{i} : S \rightarrow A_{i}$ is defined for an agent $i$ over the set of possible actions $A_{i}$. The objective of each agent $i$ is to maximize the expected discounted return $\mathbb{E}_{\pi} \left[ \sum_{t=0}^{\infty} \gamma^{t}R(s^{t}, a^{t}_{1}, a^{t}_{2}) \right]$ by following the policy $\pi$ from a given state. Therefore, the policy $\pi$ is learned by optimizing the task reward received by the agents from the environment.

\subsection{Multi-Agent Planning Problem}
A STRIPS~\citep{strips} problem is represented as $\langle P, A, I, G \rangle$ where $P$ is the set of propositions which can be used to denote facts about the world, $A$ is the set of planning actions, $I$ is the initial state, and $G$ is the goal state. Each fluent $p \in P$ is a symbolic variable that describes the current state of the environment, with each proposition representing a specific property of an object in the world.

The possible fluents for the Overcooked environment can be \textit{counter-empty} — describes whether the counter is empty or not, \textit{pot-ready} — indicates whether the soup is ready in the pot, \textit{soup-served} — indicates whether the soup has been served at the serving station, etc. $I$ denotes the propositions representing the initial state of the world, and $G$ denotes the propositions corresponding to the goal state of the world.

A planning action can be defined as $a = \langle \text{pre}(a), \text{add}(a), \text{del}(a) \rangle$ where $\text{pre}(a)$ is the set of propositions that must be true before the action can be executed, $\text{add}(a)$ are the propositions that become true after the action is performed, and $\text{del}(a)$ are the propositions that become false after the action is performed. Extending this to multiple agents, a Multi-Agent Planning task can be denoted as $\langle P, N, \{A_i\}_{i=1}^N, I, G \rangle$ where $N$ is the number of agents and $A_i$ is the set of actions for agent $i$.

We assume that agents act in parallel at each step of the plan, selecting and executing one action simultaneously. A plan is defined as a sequence of joint action sets $\left( \{a_{i}^{1}\}_{i=1}^N , \{a_{i}^{2}\}_{i=1}^N, \dots , \{a_{i}^{n}\}_{i=1}^N \right)$, where $\{a_{i}^{t}\}_{i=1}^N$ denotes the actions taken by all agents at timestep $t$,  $\{a_{i}^{t}\}$ is the action taken by the $i^{\text{th}}$ agent and $n$ is the number of steps in the plan. A plan is a solution $\Pi$ if it is a sequence of joint actions that can be applied to the initial state $I$ and results in a world state that satisfies $G$, i.e., $\Pi = \left( \{a_{i}^{1}\}_{i=1}^N , \{a_{i}^{2}\}_{i=1}^N, \dots , \{a_{i}^{n}\}_{i=1}^N \right)$ is a valid solution plan if
\begin{math}
   \{ a_{i}^{n}\}_{i=1}^N \left( \dots  \dots \left( \{ a_{i}^{2}\}_{i=1}^N \left( \{ a_{i}^{1}\}_{i=1}^N \left(I\right) \right) \right)  \right) \subseteq G.
\end{math}
\section{Interdependencies}
\subsection{Problem Statement}
We pose the human-agent teaming problem as a two-player Markov game, where the teammates act in parallel. We focus on the case where the team is trying to reach a set of goal states $S_{G}$ such that $S_{G} \subseteq S$. The states in $S_{G}$ are absorbing i.e. $\forall s \in S_{G}$ and $a^{G}_{i} \in A_{i}$, we have $T(s, \{a^{G}_{i}\}_{i=1}^{2}) = 0$. 

We represent the solution trajectory for the $i^{\text{th}}$ agent as $\tau_i = \left( a^t_i, a^{t+1}_i,\dots a^k_i \dots a^n_i\right)$ and the joint-action solution trajectory of two agents starting from timestep \textit{t} and reaching a goal state at timestep \textit{n} as $\tau = \left(\left(a^{t}_{1},a^{t}_{2}\right), \left(a^{t+1}_{1}, a^{t+1}_{2}\right) \dots \left(a^{n}_{1},a^{n}_{2}\right)  \right)$. An execution trace \textit{Tr} of a policy $\pi$ from an initial state $s^{t}$ 
as is denoted as $ \left( s^{t}, a^{t}, s^{t+1}, a^{t+1}, \dots s^{n}  \right) $, where \textit{Tr} corresponds to the state-action sequence that starts at timestep t and terminates at a goal state $s^{n} \in S_{G}$ at a timestep \textit{n}, where $a^{k}= \left(a^{k}_{1},a^{k}_{2}\right)$ and $a^{k}_{i} = \pi_{i}\left( s^{k}\right)$ for the $k^{\text{th}}$ timestep. The agents receive a task reward $R_{\text{task}}$ at the end of \textit{Tr} and $\tau$ on reaching the goal state.

\emph{Given the execution trace \textit{Tr}, the joint solution trajectory $\tau$ and only the scalar task reward $R_{\text{task}}$, there is no explicit measure of the cooperation exhibited in $\tau$. To capture the cooperative interactions arising between the teammates in $\tau$, we define the concept of interdependence in the next section.}
\subsection{Mapping the Markov Game to STRIPS}

In a Markov Game, the state at a given timestep $s_t \in S$ is typically represented as a high-dimensional vector. We can instead describe $s_t$ as a symbolic state consisting of a set of true propositions $p_t$ that denote relevant facts about the current world state. This allows us to represent each state as a finite set of meaningful symbolic facts.

Formally, there exists a function $\mathcal{F} : S \rightarrow 2^{P}$ that maps a state $s_t$ to the corresponding set of true propositions $p_t$. For example, consider Fig.~\ref{fig:coor-example}. The predicate \textit{counter-empty} denotes whether the middle counter is empty. Suppose the green-hat agent ($A_2$) takes an action to place an onion on the counter. Before this action, the proposition \textit{counter-empty} is true in state $s_t$, but after the action, it becomes false in state $s_{t+1}$. Mapping states to symbolic propositions thus enables us to capture the effects of agents' actions in terms of relevant symbols.

Recall from the execution trace \textit{Tr} of the Markov Game that at time $t$, the world is in state $s^t$. Taking joint action $a^t$ causes a transition to the next state $s^{t+1}$. Using the mapping $\mathcal{F}$, we can represent this transition symbolically as $(p_t, a^t, p_{t+1})$ where $p_t = \mathcal{F}(s^t)$ and $p_{t+1} = \mathcal{F}(s^{t+1})$. Similarly, the joint action $a^t = (a_1^t, a_2^t)$ can be mapped to a symbolic representation. For each individual action $a_i^t$, there exists a corresponding STRIPS-style planning action such that $\text{pre}(a_i^t) \subseteq p_t$, $\text{add}(a_i^t) \subseteq p_{t+1}$ and $\text{del}(a_i^t) \subseteq P \setminus p_{t+1}$.
Thus, the entire solution trajectory $\tau$ can be represented as a joint solution plan $\Pi$, where each single-agent action $a_i^t$ is expressed as $a_i^t = \langle \text{pre}(a_i^t), \text{add}(a_i^t), \text{del}(a_i^t) \rangle$.
This representation allows us to systematically track the preconditions and effects of individual agent actions in the trajectory using symbolic propositions, which in turn enables us to analyze and capture interdependencies between them.
\subsection{Agent Interdependencies}

Given a joint-action solution trajectory $\tau$ and the solution trajectory $ \tau_{i}$ for an agent i, we define the following properties about $\tau$ and $\tau_{i}$ to formalize the concept of interdependence for the solution trajectory:

\begin{definition}
\textnormal{ For $\tau$, we define \emph{Interdependence} as a pair of actions $(a_i^{t_0 + k}, a_j^{t_0})_{i \neq j}$ such that $add(a_j^{t_0}) \subseteq pre(a_i^{t_0 + k})$. An interdependent pair of actions $(a_i^{t_0 + k}, a_j^{t_0})_{i \neq j}$ has two agents, a \textit{Giver} agent performing the action $a_j^{t_0}$ and a \textit{Receiver} agent performing the action $a_i^{t_0 + k}$.  Each interdependent pair of actions is going to be associated with an object $\text{obj}_{\text{int}}$. 
}

\textnormal{We define \emph{Interdependence} as a pair of actions $(a_i^{t_0 + k}, a_j^{t_0})_{i \neq j}$ such that $add(a_j^{t_0}) \subseteq pre(a_i^{t_0 + k})$.}
\end{definition}
\begin{definition}
\textnormal{For any object in the world and a starting timestep \( t_0 \), the \emph{object influence trajectory} from time \( t_0 \), denoted by \(\tau_{\text{obj}}^{t_0}\), captures all state transitions in the plan from timestep \( t_0 \) onward where this object is involved. 
\[
\begin{aligned}
\tau_{\text{obj}}^{t_0} = \bigl\{ (p_t, a_t, p_{t+1}) \mid\ & t \geq t_0, 
 \exists p \in \text{pre}(a_t) \cup \text{add}(a_t) \cup \text{del}(a_t) \\ 
& \text{ where } \text{obj} \in \textit{O}(p) \bigr\}
\end{aligned}
\]
where \(\textit{O}(p)\) denotes the set of objects mentioned in proposition \( p \).
In other words, \(\tau_{\text{obj}}^{t_0}\) includes all transitions from timestep \( t_0 \) onward where the object explicitly appears in the action's conditions or effects. }
\end{definition}
\begin{definition} 
\textnormal{An interdependence is a \emph{Goal Reaching Interdependence} if the final state of the object associated with that interdependence ($\text{obj}_\text{int}$) is also present in the set of goal predicates.
Let $p^{\text{obj}_\text{int}}_{\text{final}}$ denote the final symbolic state of the object associated with the interdependence, corresponding to the last entry in $\tau_{\text{obj}_\text{int}}^{t_0}$. Then, the interdependence Int is classified as a Goal Reaching Interdependence if:
\[
p^{\text{obj}_\text{int}}_{\text{final}} \subseteq p^G_n,
\]
where $p^G_n$ is the set of goal propositions at timestep $n$.}
\end{definition}
\begin{definition}
\textnormal{
Let \( p^{\text{obj}}_t \) denote the predicate for that object at timestep \( t \), therefore containing information about the state of \( \text{obj} \) at \( t \). An interdependence 
\(
\text{Int} = (a_i^{t_0 + k}, a_j^{t_0})
\) 
associated with object 
\(
\text{obj}_{\text{int}}
\) 
is said to be a \textit{Non-looping Interdependence} if the following conditions hold:
\begin{enumerate}
    \item The \textit{giver agent} (agent \( j \)), who gives the object \( \text{obj}_{\text{int}} \) at timestep \( t_0 \), \textit{does not receive} the object back in the same state at any future timestep \( t > t_0 + k \):
    \begin{align*}
\nexists t > t_0 + k, \quad \text{s.t.} \quad 
& \text{agent } j \text{ receives } \text{obj}_{\text{int}} \\
& \text{in the same state as at time } t_0
\end{align*}
    \item The \textit{receiver agent} (agent \( i \)), who receives the object at timestep \( t_0 + k \), \textit{did not have} the object in that same state at any time \( t < t_0 + k \):
    \[
    \nexists t < t_0 + k, \quad \text{s.t.} \quad \text{agent } i \text{ had } \text{obj}_{\text{int}} \text{ in the same state}
    \]
\end{enumerate}
}
\end{definition}

\begin{definition}
\textnormal{
An interdependence \( \texttt{Int}= (a^k_i, a^{k-t}_j) \) is a \textbf{Constructive Interdependence}, if it is a \emph{Goal Reaching Interdependence} and a \emph{Non-looping Interdependence}.
}
\end{definition}
Consider a scenario in the Counter Circuit layout where agent \( j \) places an onion on the counter at timestep \( t_0 \) via action \( a_j^{t_0} \), whose effect is \( add(a_j^{t_0}) = \{\texttt{onion-on-counter}\} \). Subsequently, at timestep \( t_0 + k \), agent \( i \) performs action \( a_i^{t_0 + k} \) to pick up the onion from the counter, with precondition \( pre(a_i^{t_0 + k}) = \{\texttt{onion-on-counter}\} \). This pair of actions \( (a_i^{t_0 + k}, a_j^{t_0}) \) constitutes a \textit{sequential interdependence} \(\text{Int}\) linked to the object \( \text{obj}_{\text{int}} = \text{onion} \). The associated object influence trajectory \( \tau^{t_0}_{\text{onion}} \) captures all state transitions involving the onion, culminating in a final state where the soup contains the onion. Provided that the onion is not returned to agent \( j \) in the same state and that agent \( i \) had not previously held the onion in that state, this interdependence is \textit{Non-looping}. Consequently, this interaction qualifies as a Constructive Interdependence. A \textit{Trigger} action for an agent is placing the onion on the counter, since it could potentially be the precondition for the other agent picking that onion from the counter. 
\begin{figure}[htbp]
  \centering
  \includegraphics[width=0.45\textwidth]{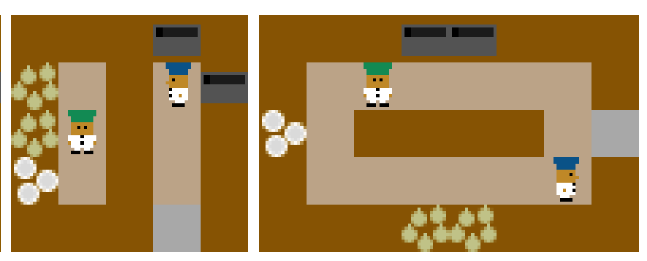}
  \caption{Left: Forced coordination layout which is a required cooperation (RC) setting. Right: Counter circuit layout which is an non-required cooperation (Non-RC) setting.}
  \label{fig:your-label}
\end{figure}
\section{Experiments}
We evaluate state-of-the-art (SOTA) zero-shot cooperation (ZSC) agents in human-AI teams using the proposed metric within a benchmark environment. Our analysis focuses on these questions:
\begin{itemize}[leftmargin=*]
\item \textit{RQ1}: Can ZSC agents recognize and adapt to partners who initiate cooperative strategies?
\item \textit{RQ2}: To what extent do teams with ZSC agents and human partners exhibit cooperative behavior in Non-Required Cooperation (Non-RC) scenarios?
\item \textit{RQ3}: How does cooperative behavior differ between RC and Non-RC settings when ZSC agents team with humans?
\end{itemize}

\textbf{Metrics}:
We quantify cooperation in teams using two key metrics: constructive interdependencies($Int_{\text{cons}}$) which capture instances where agents’ actions meaningfully contribute to task completion through interdependent actions and non-constructive interdependencies($Int_{\text{non-cons}}$) which quantifies unproductive or redundant interdependencies between agents’ actions. We also measure how many interdependencies are initiated by the teammates($\%P^{\text{trig}}_{\text{tot-sub}}$), and of those, how many are \emph{not accepted and acted upon} by the ZSC agents($\%P^{\text{not-trig-acc}}_{\text{trig}}$). 

\textbf{Environment: }The environment is a fully observable, timed gridworld where a team of two players must collaboratively prepare and deliver three soups within a fixed time limit. It contains onion dispensers, dish dispensers, pots, serving stations, and empty counters. Players act concurrently at each timestep and can either move or interact with objects. To complete a soup and get a reward, the team must collect and place three onions in a pot, wait for it to cook, retrieve a dish, transfer the soup, and deliver it. Each player and counter can hold only one object at a time. Task rewards are shared between both players upon successful delivery, which is \textit{expected to incentivize efficient collaboration}. 
\newline
\textbf{SOTA Methods}: FCP~\citep{fcp}, MEP~\citep{mep}, HSP~\citep{hsp}, and COLE~\citep{cole} utilize a two-stage training pipeline. First, a diverse population of partners is generated through self-play. Next, an ego agent is trained via reinforcement learning by interacting with sampled partners from this population, optimizing primarily for episodic task reward. This task reward, which serves as the training signal, is also the main evaluation metric used to assess cooperation when these agents are paired with previously unseen teammates, including humans.
\newline
\textbf{Evaluation Partners: } Using the proposed metric, we analyze the cooperative behavior of ZSC agents, on the forced coordination (RC) and counter circuit (non-RC) layout (Fig.~\ref{fig:your-label}), in the following settings:
\begin{enumerate}[noitemsep]
    \item \textit{Scripted Cooperative Agent:} Agents are paired with a scripted partner executing a fixed cooperative strategy. This setting tests whether ZSC agents can recognize and adapt to consistently cooperative behaviors (\textit{RQ1}).
    \item \textit{Self-Play:} Agents are paired with an identical copy of themselves. Serving as a baseline, this evaluates their ability to coordinate when interacting with the same policy (\textit{RQ2}).
    \item $H_{\text{proxy}}$: Agents interact with learned human behavior models~\citep{zsceval}. Since scaling user studies is costly, this provides an additional proxy for assessing agent cooperation with humans.(\textit{RQ2, RQ3})
    \item \textit{Human Teammates:} We conduct a user study pairing ZSC agents with 36 human participants to measure cooperative behavior in real-world human-agent teams (\textit{RQ1,RQ2, RQ3}).
\end{enumerate}
\textbf{User Study:} We recruited 36 participants from our university in the range from 18 to 31 who were pursuing either an undergraduate or a graduate degree. We conducted a pilot study on 5 participants spread across each of the two evaluation domains. The final study had a sample size of 31 participants. Participants had an average age of 20.75 years, and a median age of 22.5 years. Out of the 36, there were 24 male participants and 12 female participants. 23 participants (63.9\%) reported to not have any familiarity with playing the Overcooked game earlier, and the remaining 13 (36.1\%) were familiar with the game. 
\section{Results}
In this section, we present the evaluation of ZSC agents paired with various partners. Using our proposed metric, we quantify the cooperative behavior exhibited by these teams across different settings.
\begin{table}[ht]
\centering
\setlength{\tabcolsep}{3pt}
\renewcommand{\arraystretch}{1.1}
\begin{subtable}{\linewidth}
\centering
{\fontsize{10.5}{9}\selectfont
\setlength{\tabcolsep}{3pt}
\begin{tabular}{@{}lcccc@{}}
\toprule[1.5pt]
\textbf{Agent} 
& \textbf{Task Reward} 
& \textbf{$\text{Int}_{\text{cons}}$} 
& \textbf{$\text{Int}_{\text{non-cons}}$} 
& \textbf{$\text{Int}_{\text{cons}}^{\text{coop}}$} \\
\midrule[1pt]
COLE & 36 & 0.60 & 1.20 & 4.425 \\
\midrule
MEP & \cellcolor{blue!10}43.33 & \cellcolor{blue!10}0.83 & 1.67 & 6.59 \\
\midrule
HSP & 0 & 0 & 0.50 & 0 \\
\midrule
FCP & 0 & 0 & 0.17 & 0 \\
\bottomrule[1.5pt]
\end{tabular}
}
\caption{Average rewards and cooperation metrics across 50 runs. $\text{Int}_{\text{cons}}^{\text{coop}}$ represents the interdependence if both agents followed the coordination policy.}
\label{sub:scripted-rew}
\end{subtable}
\vspace{6pt}
\begin{subtable}{\linewidth}
\centering
\begin{tabular}{@{}ccc@{}}
\toprule[1.5pt]
\textbf{Agent} & \boldmath$\%\!P^{\text{trig}}_{\text{tot-sub}}$ & \boldmath$\%\!P^{\text{not-trig-acc}}_{\text{trig}}$ \\
\midrule[1pt]
COLE & \cellcolor{red!10}58.5 & \cellcolor{red!10}88.88 \\
\midrule
MEP & 41.28 & 75.55 \\
\midrule
HSP & \cellcolor{red!10}58.57 & \cellcolor{red!10}100.0 \\
\midrule
FCP & 41.79 & 100.0 \\
\bottomrule[1pt]
\end{tabular}
\caption{Average response rates of ZSC agents to coordination attempts. $\%P^{\text{trig}}_{\text{tot-sub}}$ shows the percentage of interdependencies triggered by the scripted agent, while $\%P^{\text{not-trig-acc}}_{\text{trig}}$ indicates the percentage of triggered interdependencies that were not accepted by the ZSC agents.}
\label{sub:scripted-response}
\end{subtable}
\caption{Analysis of ZSC agents' performance when paired with a cooperative scripted partner in Non-RC layout. \cellcolor{blue!10}Blue cells highlight cases with high task reward despite much lower cooperation than the joint coordination policy; \cellcolor{red!10}red cells mark frequently triggered interdependencies by human which are frequently rejected by ZSC agents.
}
\end{table}
\subsection{ZSC Agent paired with Cooperative Partner}
We assess whether ZSC agents can recognize and adapt to a partner employing a known coordination policy by quantifying the resulting coordination using our proposed metric. Specifically, we evaluate teams in the Counter Circuit layout, implementing the coordination strategy introduced by \citet{carroll}, as illustrated in Fig.~\ref{fig:coor-example}. In this strategy, the green-hat chef places onions on the counter, while the blue-hat chef picks them up, places them in the pot, and later serves the soup. Our experimental setup assigns the scripted agent to the green-hat role and the ZSC agent to the blue-hat role, testing the agent’s ability to complement and adapt to its partner’s cooperative behavior.
\newline
The joint coordination policy exhibits sequential interdependence, which we capture using the constructive interdependency metric ($Int_{\text{cons}}$). We also measure constructive interdependencies when both agents strictly follow the joint coordination policy ($Int_{\text{cons}}^{\text{coop}}$). This represents the upper bound on coordination achievable if the ZSC agent fully adapts to the scripted partner. Comparing these values enables us to evaluate the extent to which ZSC agents successfully recognize and complement cooperative strategies in a Non-RC setting.
\newline
From Table~\ref{sub:scripted-rew}, we observe that all ZSC agents exhibit low levels of constructive interdependence when paired with a scripted partner following a coordination policy. For example, COLE achieves a task reward of 36 but only reaches $Int_{\text{cons}} = 0.60$, far below the cooperative upper bound of $Int_{\text{cons}}^{\text{coop}} = 4.425$. Similarly, MEP obtains the highest reward of 43.33 among the agents but still falls short in coordination, with $Int_{\text{cons}} = 1.83$ compared to $Int_{\text{cons}}^{\text{coop}} = 6.59$. As shown in Table~\ref{sub:scripted-response}, the scripted agent frequently attempts to initiate interdependencies ($\%P^{\text{trig}}_{\text{tot-sub}}$ of 58.5 for COLE and 41.28 for MEP), yet these are often ignored by the ZSC agents, with rejection rates of 88.88\% and 75.55\% respectively. Despite the partner’s consistent cooperative behavior, HSP and FCP entirely fail to coordinate, with zero constructive interdependence and 100\% rejection rates.
\subsection{Task vs Teaming Performance of ZSC in Self-Play}
We evaluate whether agents can induce cooperative strategies when paired with identical copies of themselves (Table~\ref{table:best-self}). We find that while task rewards are generally high, the level of constructive interdependence varies widely. For example, COLE and HSP both achieve a reward of 120 in non-RC, but differ substantially in $Int_{\text{cons}}$ (10.23 vs. 1.82), indicating that similar task outcomes can arise from vastly different levels of coordination. In RC layouts, however, we observe a strong alignment between task reward and $Int_{\text{cons}}$ (e.g., COLE: 200 and 30.43; MEP: 140 and 29.5).  Furthermore, across all non-RC cases, $Int_{\text{non-cons}}$ consistently exceeds $Int_{\text{cons}}$, as seen in MEP (10.19 vs. 1.05) and HSP (12.32 vs. 1.82), indicating that when interdependencies do arise, they are often unproductive or misaligned.
\subsection{Task vs Teaming Performance of ZSC Agents Paired with $H_{\text{proxy}}$}
As observed in Table~\ref{table:best-proxy}, constructive interdependencies remain low in Non-RC layout, indicating weak cooperation when ZSC agents are paired with $H_{\text{proxy}}$. For example, MEP achieves the highest task reward in Non-RC (140) but exhibits the lowest $\text{Int}_{\text{cons}}$ (1.05), highlighting a decoupling between reward and cooperative behavior. By contrast, in RC layout, we observe a marked improvement in constructive interdependencies. COLE, for instance, attains both the highest task reward (200) and the highest $\text{Int}_{\text{cons}}$ (35), demonstrating strong alignment between cooperation and performance. This pattern is consistent across agents—highlighting the inadequacy of task reward as a proxy for teamwork in Non-RC settings.
\begin{table}[]
\centering
{\fontsize{10.5}{9}\selectfont
\setlength{\tabcolsep}{3pt}
\begin{tabular}{@{}lcccccc@{}}
\toprule[1.5pt]
\textbf{Agent} 
& \multicolumn{2}{c}{\textbf{Task Reward}} 
& \multicolumn{2}{c}{\textbf{$\text{Int}_{\text{cons}}$}} 
& \multicolumn{2}{c}{\textbf{$\text{Int}_{\text{non-cons}}$}} \\
\cmidrule[1pt]{2-7}
& \text{Non-RC} & \text{RC} & \text{Non-RC} & \text{RC} & \text{Non-RC} & \text{RC} \\
\midrule[1pt]
COLE & 120 & \cellcolor{green!15}200 & 10.23 & \cellcolor{green!15}30.43 & 6.58 & 6.83 \\
\midrule
MEP & \cellcolor{blue!10}100.00 & 140 & \cellcolor{blue!10}1.05 & 29.5 & \cellcolor{red!10}10.19 & 7.53 \\
\midrule
HSP & 120 & 100 & 1.82 & 22.87 & \cellcolor{red!10}12.32 & 10.667 \\
\midrule
FCP & 60 & 80 & 0 & 16.0 & 0 & 16.0 \\
\bottomrule[1.5pt]
\end{tabular}
}
\caption{Average rewards and cooperation metrics of ZSC agents in self-play across Non-RC and RC layouts. 
\cellcolor{green!15}Green cells highlight correlation of reward and cooperation in RC; \cellcolor{blue!10}blue cells indicate high task reward despite low cooperation in Non-RC; \cellcolor{red!10}red cells mark non-constructive interdependencies.}
\label{table:best-self}
\end{table}
\begin{table}[]
\centering
{\fontsize{10.5}{9}\selectfont
\setlength{\tabcolsep}{3pt}
\begin{tabular}{@{}lcccccc@{}}
\toprule[1.5pt]
\textbf{Agent} 
& \multicolumn{2}{c}{\textbf{Task Reward}} 
& \multicolumn{2}{c}{\textbf{$\text{Int}_{\text{cons}}$}} 
& \multicolumn{2}{c}{\textbf{$\text{Int}_{\text{non-cons}}$}} \\
\cmidrule[1pt]{2-7}
& \text{Non-RC} & \text{RC} & \text{Non-RC} & \text{RC} & \text{Non-RC} & \text{RC} \\
\midrule[1pt]
COLE & 100 & \cellcolor{green!15}200 & 7.52 & \cellcolor{green!15}35.00 & 10.98 & 3.32 \\
\midrule
MEP & \cellcolor{blue!10}140.00 & 180 & \cellcolor{blue!10}1.05 & 29.5 & 12.21 & 9.53 \\
\midrule
HSP & \cellcolor{blue!10}160 & 180 & \cellcolor{blue!10}4.5 & 30.00 & \cellcolor{red!10}11.32 & 7.667 \\
\midrule
FCP & 100 & 120 & 6.021 & 24 & \cellcolor{red!10}15.02 & 4.0 \\
\bottomrule[1.5pt]
\end{tabular}
}
\caption{Average rewards and cooperation metrics for ZSC agents when paired with $H_{\text{proxy}}$. \cellcolor{green!15}Green cells highlight correlation of reward and cooperation in RC; \cellcolor{blue!10}blue cells indicate high task reward despite low cooperation in Non-RC; \cellcolor{red!10}red cells mark non-constructive interdependencies.}
\label{table:best-proxy}
\end{table}
\begin{table}[ht!]
\centering
{\fontsize{10.5}{9}\selectfont
\setlength{\tabcolsep}{4pt}
\begin{subtable}{0.95\columnwidth}
\centering
\begin{tabular}{@{}lcccccc@{}}
\toprule[1.5pt]
\textbf{Agent} 
& \multicolumn{2}{c}{\textbf{Task Reward}} 
& \multicolumn{2}{c}{\textbf{$\text{Int}_{\text{cons}}$}} 
& \multicolumn{2}{c}{\textbf{$\text{Int}_{\text{non-cons}}$}} \\
\cmidrule[1pt]{2-7}
& \text{Non-RC} & \text{RC} & \text{Non-RC} & \text{RC} & \text{Non-RC} & \text{RC} \\
\midrule[1pt]
COLE & \cellcolor{blue!10}76.21 & 56.87 & \cellcolor{blue!10}1.89 & 11.37 & \cellcolor{yellow!20}6.29 & 2.87 \\
\midrule
MEP & \cellcolor{blue!10}50 & 44.1 & \cellcolor{blue!10}0.92 & 8.69 & 1.28 & 2.76 \\
\midrule
HSP & 41.11 & \cellcolor{green!15}60.55 & \cellcolor{red!10}1.38 & \cellcolor{green!15}12.05 & 2.13 & 3.08 \\
\midrule
FCP & 22.55 & 35.34 & 0.97 & 7.06 & 0.872 & 3.44 \\
\bottomrule[1.5pt]
\end{tabular}
\caption{Average rewards and cooperation metrics for ZSC agents across all runs with human participants.}
\label{table:merged-hum-1}
\end{subtable}

\vspace{4mm}

\begin{subtable}{0.95\columnwidth}
\centering
\begin{tabular}{@{}lcccccc@{}}
\toprule[1.5pt]
\textbf{Agent} 
& \multicolumn{2}{c}{\textbf{Task Reward}} 
& \multicolumn{2}{c}{\textbf{$\text{Int}_{\text{cons}}$}} 
& \multicolumn{2}{c}{\textbf{$\text{Int}_{\text{non-cons}}$}} \\
\cmidrule[1pt]{2-7}
& \text{Non-RC} & \text{RC} & \text{Non-RC} & \text{RC} & \text{Non-RC} & \text{RC} \\
\midrule[1pt]
COLE & 120 & 100 & 3.79 & 20.00 & 3.0 & 0.33 \\
\midrule
MEP & \cellcolor{blue!10}80.00 & 120 & \cellcolor{blue!10}1.285 & 24.0 & \cellcolor{yellow!20}7.47 & 0 \\
\midrule
HSP & 80 & \cellcolor{green!15}120 & 3.5 & \cellcolor{green!15}25.00 & 1.25 & 1.667 \\
\midrule
FCP & 60 & 100 & 3.667 & 20 & 1.334 & 3.0 \\
\bottomrule[1.5pt]
\end{tabular}
\caption{Rewards and cooperation metrics for top-performing human-agent teams for each ZSC agent.}
\label{table:merged-hum-best-1}
\end{subtable}

\vspace{4mm}

\begin{subtable}{1\columnwidth}
\centering
\begin{tabular}{@{}lcccc@{}}
\toprule[1.5pt]
\textbf{Agent} 
& \multicolumn{2}{c}{\textbf{\%P$^{\mathrm{trig}}_{\mathrm{tot-sub}}$}}
& \multicolumn{2}{c}{\textbf{\%P$^{\mathrm{not\,trig\,-acc}}_{\mathrm{trig}}$}} \\
\cmidrule[1pt]{2-5}
& \text{Non-RC} & \text{RC} & \text{Non-RC} & \text{RC} \\
\midrule[1pt]
COLE & 60.28 & 45.28 & 70.05 & 38.34 \\
\midrule
MEP & \cellcolor{red!10}66.82 & 43.57 & \cellcolor{red!10}82.39 & 39.82 \\
\midrule
HSP & 52.22 & 42.92 & 80.85 & 40.58 \\
\midrule
FCP & \cellcolor{red!10}58.30 & 43.62 & \cellcolor{red!10}98.41 & 36.84 \\
\bottomrule[1.5pt]
\end{tabular}
\caption{Response rates of ZSC agents to coordination attempts by human teammates. $\%P^{\text{trig}}_{\text{tot-sub}}$ shows the percentage of interdependencies triggered by the humans, while $\%P^{\text{not-trig-acc}}_{\text{trig}}$ indicates the percentage of triggered interdependencies that were not accepted by the ZSC agents.}
\label{table:merged-hum-2}
\end{subtable}

\caption{Comprehensive analysis of human-agent team performance across Non-RC and RC settings. \cellcolor{green!15}Green cells highlight correlation of reward and cooperation in RC; \cellcolor{blue!10}blue cells indicate high task reward despite low cooperation in Non-RC; \cellcolor{red!10}red cells mark frequently triggered interdependencies by human which are frequently rejected by ZSC agents;\cellcolor{red!10}yellow cells mark non-constructive interdependencies.}
\label{table:merged-hum-all}
}
\end{table}
\subsection{Task vs Teaming Performance of ZSC Agents paired with Human Teammates}
From Table~\ref{table:merged-hum-1}, constructive interdependencies ($\text{Int}_{\text{cons}}$) remain consistently low in Non-RC layouts even with human partners. For example, COLE and HSP achieve reasonably high task rewards (76.21 and 41.11) but low $\text{Int}_{\text{cons}}$ (1.89 and 1.39 respectively). This trend persists among the best-performing teams (Table~\ref{table:merged-hum-best-1}), where MEP achieves a reward of 80 but with only 1.285 constructive interdependencies, underscoring the limited cooperation despite successful task completion.  
In the RC layout, we observe a strong positive correlation between task performance and constructive interdependencies. For instance, the HSP agent achieves the highest task reward of 120 when paired with a human participant, and this is accompanied by the highest number of constructive interdependencies (25.00) as shown in Table~\ref{table:merged-hum-best-1}. Thus, task reward can serve as a reliable proxy for cooperative behavior in RC settings, in contrast to non-RC layouts where such alignment breaks down. Table~\ref{table:merged-hum-2} shows that humans frequently initiate cooperative interactions (e.g., MEP triggers 66.82\% interdependencies in Non-RC), but ZSC agents often reject these attempts. We also found a strong positive correlation in RC settings (Pearson $r = 0.81$, $p < 0.001$), and weak correlation in Non-RC settings, (Pearson $r = 0.19$), confirming that agents can achieve high reward without meaningful cooperation. \\
Using the proposed metric, our analysis of human-agent teams reveals the following key insights:
\begin{itemize}[nosep]
    \item When paired with partners—scripted or human—who attempt to initiate cooperation, ZSC agents fail to adapt to their partners and  frequently reject coordination attempts.
    \item In Non-RC layouts, ZSC agents can achieve high task performance, but they do not induce cooperative behavior. Agents that perform well demonstrate significantly low levels of constructive interdependence. 
    \item In RC layouts, task reward aligns much more strongly with inter-agent coordination. Agents that perform well also demonstrate higher levels of constructive interdependence.
    \item In cases where interdependence emerges in Non-RC settings, these are frequently non-constructive, indicating misaligned or ineffective coordination.
\end{itemize}
These results reveal the inadequacy of task reward as a standalone metric for evaluating cooperation in human-agent teams in Non-RC settings. This work highlights a critical gap in current state-of-the-art for Zero-Shot Coordination: their limited ability to engage in meaningful cooperation when paired with partners attempting to coordinate. 
\subsection{Incorporating Teaming Reward into Training}
\begin{figure*}[ht!]
  \centering
  \begin{subfigure}[b]{0.48\linewidth}
    \centering
    \includegraphics[width=\linewidth]{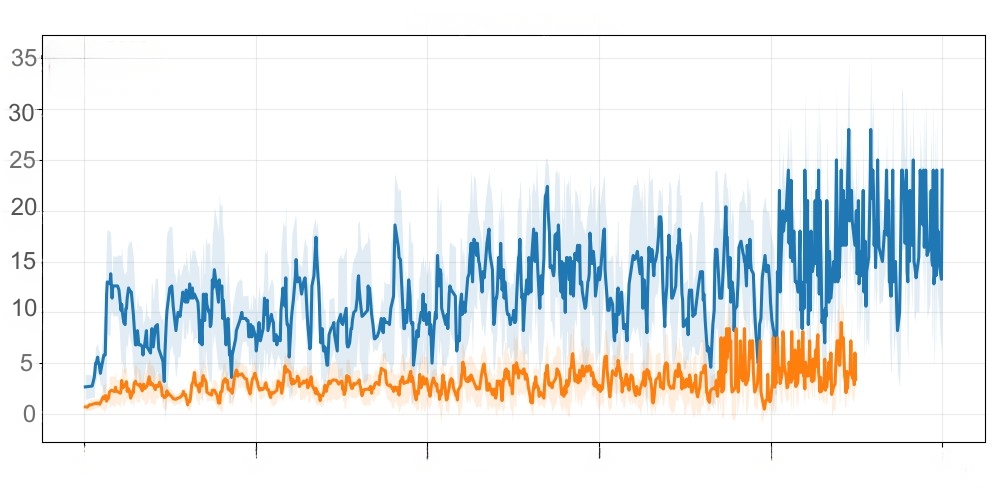}
    \caption{FCP: Teaming Reward}
    \label{fig:fcp-team}
  \end{subfigure}
  \hfill
  \begin{subfigure}[b]{0.48\linewidth}
    \centering
    \includegraphics[width=\linewidth]{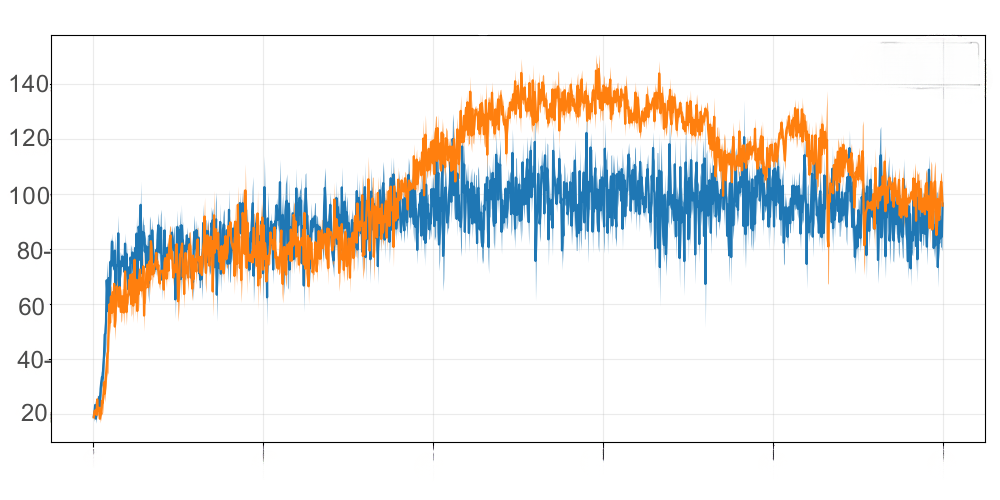}
    \caption{FCP: Task Reward}
    \label{fig:fcp-task}
  \end{subfigure}

  \vspace{1.5ex}

  \begin{subfigure}[b]{0.48\linewidth}
    \centering
    \includegraphics[width=\linewidth]{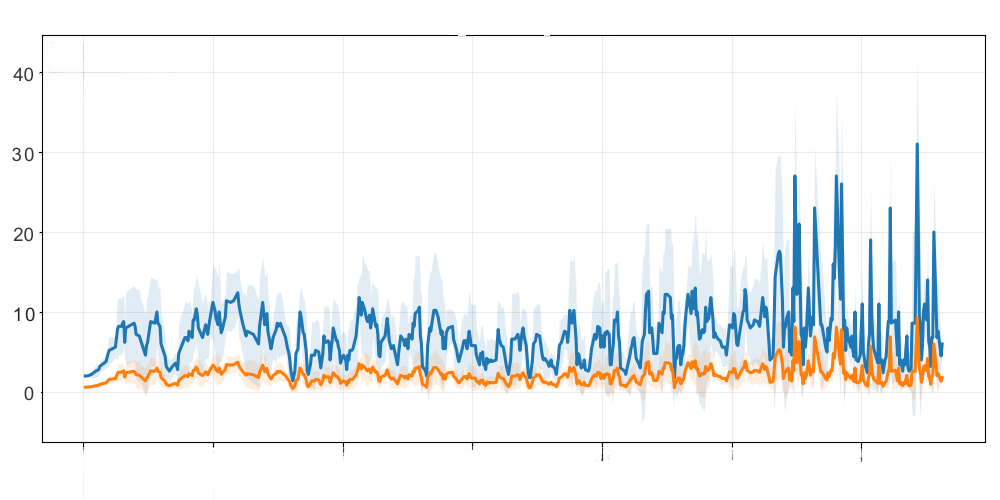}
    \caption{MEP: Teaming Reward}
    \label{fig:mep-team}
  \end{subfigure}
  \hfill
  \begin{subfigure}[b]{0.48\linewidth}
    \centering
    \includegraphics[width=\linewidth]{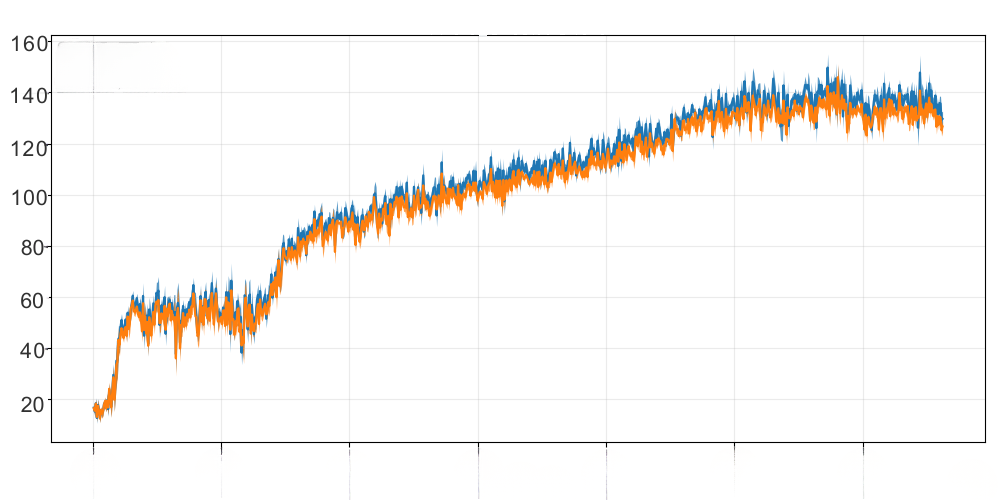}
    \caption{MEP: Task Reward}
    \label{fig:mep-task}
  \end{subfigure}

  \vspace{1.5ex}

  \begin{subfigure}[b]{0.48\linewidth}
    \centering
    \includegraphics[width=\linewidth]{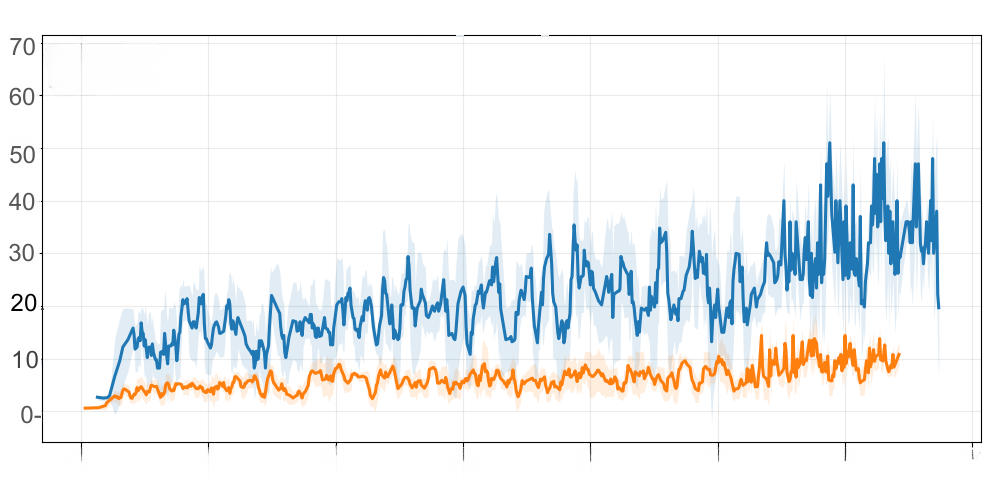}
    \caption{HSP: Teaming Reward}
    \label{fig:hsp-team}
  \end{subfigure}
  \hfill
  \begin{subfigure}[b]{0.48\linewidth}
    \centering
    \includegraphics[width=\linewidth]{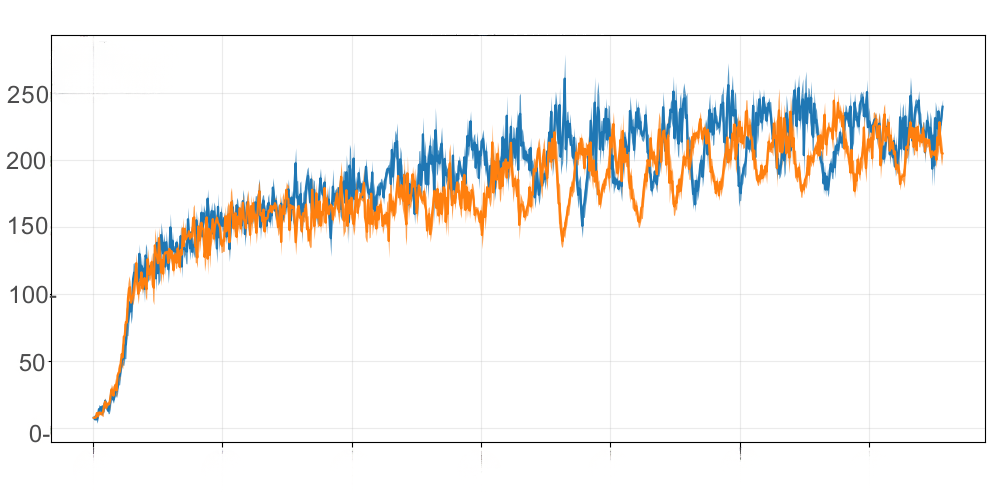}
    \caption{HSP: Task Reward}
    \label{fig:hsp-task}
  \end{subfigure}

  \caption{Training curves comparing agents trained with the original task-only reward (orange curves) and the modified reward (blue curves) that incorporates interdependence ($\alpha$ = 0.3). All agents are trained for 5 million timesteps. The orange curves correspond to training with $\textbf{r}_{original}$, while the blue curves correspond to training with $\textbf{r}_{modified}$. Across all three algorithms (FCP, MEP, HSP), adding the teaming reward leads to a consistent increase in construcive interdependencies, while task performance remains unchanged.  }
  \label{fig:team_task_grid}
\end{figure*}

Motivated by this gap, we conducted preliminary experiments incorporating the proposed cooperation metric into the training of SOTA algorithms. Concretely, we modify the training reward as follows, where $r_{task}$ is the task reward and $r_{teaming}$ is the teaming reward measured by constructive interdependencies: $\textbf{r}_{modified} = r_{task} + \alpha \times r_{teaming}, \textbf{r}_{original} = r_{task}$.  \\
We train SOTA ZSC agents under both reward formulations and compare their learning dynamics. From Fig.~\ref{fig:team_task_grid}, we observe that  agents trained with the modified reward consistently increase their teaming reward throughout training, indicating that they learn to engage in behaviors that lead to constructive interdependencies. In contrast, agents trained with the original reward show nearly stagnant teaming reward, confirming that standard training does not incentivize cooperative interactions. Importantly, we observe that task performance remains comparable between the modified and original training rewards, suggesting that incorporating interdependence does not significantly compromise task reward. Overall, these findings show that incorporating the proposed cooperation metric directly into the learning objective enables SOTA ZSC agents to learn cooperative interactions without compromising on the task reward, addressing a key limitation identified in our evaluation.
\section{Conclusion}
In this work, we propose a metric to evaluate cooperation in human-AI teams, addressing key limitations of task-reward-based evaluations in zero-shot cooperation (ZSC). 
Our findings reveal that current state-of-the-art ZSC agents often achieve high task rewards without engaging in meaningful coordination with their partners. Through controlled evaluation with a scripted agent, human proxy, and real human participants, we showed that these agents fail to recognize or respond appropriately to cooperative signals, particularly in environments where cooperation is optional. This highlights that task reward alone is an insufficient proxy for cooperative behavior in scenarios where cooperation is optional, emphasizing the need of explicit metrics to accurately quantify inter-agent coordination. 
Future work would include broadening this metric to include other kinds of interdependencies and cooperative behaviors. Another research direction is to use the interdependence metric as an additional reward signal to guide learning towards effective cooperation, rather than relying on coordination to emerge implicitly from the task reward. In non-RC settings, the \emph{shadowed equilibrium problem}~\citep{matignon, fulda} causes agents to not explore the cooperative strategies during training, since multiple equilibria exist including the non-cooperative strategies. We also provide preliminary experiments showing that integrating the interdependence metric as a reward signal encourages agents to actively recognize and pursue coordination during exploration, potentially learning to play with a diverse set of partners and reducing miscoordination in human-agent teaming scenarios.
Ultimately, this work paves the way for developing ZSC agents that not only succeed at task performance but also robustly cooperate with diverse human behaviors.
\section{Acknowledgments}
This research is supported primarily by ONR grant N0001423-1-2409. This research is also supported by DARPA grant HR00112520016, and gifts from Qualcomm and Amazon. Special thanks to Dhanush Giriyan for guidance and technical support

\bibliography{aaai2026}
\section*{Appendix}
\appendix
\section{Environment Details}
At each step, players can perform either of these eight actions: stay in the same cell, move one cell up, move one cell down, move one cell to the right, move one cell to the left and interact with the object in front. The result of this action depends on the item the player is holding (empty, onion, empty dish, filled dish) and the type of object they are facing(dispenser, pot, empty counter, serving station). 
Since the environment we are working with has a distinct \textsc{\textit{interact}} action, we can enumerate all possible outcomes of the the interact action, and use these as our sub-tasks - Pick up onion from onion dispenser,  Pick up onion from counter, Pick up dish from dish dispenser, Pick up dish from counter, Place onion in pot, Place onion on counter, Get soup from pot, Place dish on counter, Get soup from pot, Place soup on counter, Serve soup in serving station. 
\section{Pipeline}
To support reproducibility and generalizability of our proposed cooperation metric, we provide a domain-agnostic software package\footnote{Repository: \url{https://anonymous.4open.science/r/aaai-26/}} that allows researchers to apply our analysis across any multi-agent domains. While the main paper demonstrates the utility of the metric in the Overcooked environment, the framework is explicitly designed to be decoupled from any domain-specific assumptions. The system is structured into two independent modules (which are described in detail in the next two sub-sections):

\textbf{(1) Mapping Module:} This module abstracts execution traces into a symbolic representation, generating the grounded trajectory. Given a trajectory $\tau = {(s^t, a^t, s^{t+1})}_{t=0}^{T}$ from any Markov Game, the module uses a user-defined mapping function $\mathcal{F} : S \rightarrow 2^P$ to convert each low-level state $s^t$ into a set of true symbolic propositions $p_t \subseteq P$, where $P$ is the set of domain predicates. Likewise, each agent action $a_i^t$ is mapped into a STRIPS-style operator $\langle \text{pre}(a_i^t), \text{add}(a_i^t), \text{del}(a_i^t) \rangle$, derived from the symbolic state transitions $(p_t, p{t+1})$. The mapping configuration—defining predicates, object types, and effect extraction functions—is modular and can be specified declaratively for any domain.\\
\textbf{(2) Analysis Module:} This module performs an interdependence analysis on the grounded trajectory by examining how the effects of one agent's action satisfy the preconditions of subsequent actions by teammates. The analysis module classifies such interactions into constructive (task-contributing) and non-constructive (redundant or not task-contributing) interdependencies. This module generates the count of each type of interdependence in the team's action trajectory in one round of the game. 
\begin{figure}[h]
    \resizebox{0.95\linewidth}{!}{
    \begin{tikzpicture}[
        module/.style={rectangle, draw, rounded corners, thick, text centered, minimum height=2.6em, minimum width=6.2cm, fill=blue!10},
        io/.style={rectangle, draw, thick, text centered, minimum height=2.4em, minimum width=5.8cm, fill=gray!10},
        arrow/.style={->, thick, >=Latex},
        note/.style={font=\footnotesize, align=left}
      ]

      \node[io] (input) {Raw Trajectory : \\ $(s^t, a^t, s^{t+1})_{t=0}^{T}$};
      \node[module, below=1.1cm of input] (mapping) {Mapping Module \\};
      \node[io, below=1.1cm of mapping] (grounded) {Grounded Symbolic Trajectory : \\ $(p^t, \langle \text{pre}, \text{add}, \text{del} \rangle)_{t=0}^{T}$};
      \node[module, below=1.1cm of grounded] (analysis) {Analysis Module \\ };
      \node[io, below=1.1cm of analysis] (output) {Output Metrics : Constructive / Non-constructive Interdependencies};

      \draw[arrow] (input) -- (mapping);
      \draw[arrow] (mapping) -- (grounded);
      \draw[arrow] (grounded) -- (analysis);
      \draw[arrow] (analysis) -- (output);

      \node[note, right=1.2cm of mapping] (mapnote) {
        \textbf{Requires:} \\
        - Domain predicates $P$ \\
        - Object types \\
        - Mapping function $\mathcal{F}: S \to 2^P$ \\
      };

      \node[note, right=1.2cm of analysis] (analnote) {
        \textbf{Detect sequential interdependencies:} \\
        - Identify when $a_i^t$'s effects \\
        \hspace{0.3cm} satisfy $a_j^{t+k}$'s preconditions \\
        - Categorize as: \\
        \hspace{0.3cm} Constructive (goal-relevant) \\
        \hspace{0.3cm} Non-constructive (redundant)
      };

    \end{tikzpicture}}
    \caption{Software architecture for our domain-agnostic cooperation analysis framework. The Mapping Module converts raw trajectories to symbolic STRIPS-style traces, and the Analysis Module identifies interdependencies}
    \label{fig:software-architecture}
\end{figure}

\subsection{Mapping Module}
\label{section:mapping}
The \texttt{mapping} module provides a general-purpose utility to convert trajectories from any Markov Game environment into a symbolic STRIPS-like planning formalism expressed in PDDL. This abstraction is achieved by defining a declarative mapping between environment states and a set of domain-specific predicates that describe the symbolic state of the world.

The module is designed to be domain-agnostic. Users define a configuration file specifying:
\begin{itemize}
    \item The list of symbolic predicates relevant to their environment.
    \item Custom extraction functions for identifying which predicates hold in a given state.
    \item Mappings from low-level environment actions to high-level symbolic actions, including their preconditions, add effects, and delete effects.
\end{itemize}

Given a trajectory consisting of $(s^t, a^t, s^{t+1})$ tuples, the mapping module automatically generates:
\begin{itemize}
    \item A symbolic trace of world states $p_t = \mathcal{F}(s^t)$.
    \item A sequence of STRIPS-style operator instances for each agent's action, of the form:
    \[
    a_i^t = \langle \text{pre}(a_i^t), \text{add}(a_i^t), \text{del}(a_i^t) \rangle.
    \]
\end{itemize}

The output is a valid, grounded PDDL trace. Internally, the codebase is modular and allows plugging in new domain environments with minimal changes — only the symbolic interface for states and actions needs to be defined. This module supports multi-agent turn-based trajectories by assuming alternating agent moves and handles each agent's action separately when computing symbolic transitions. Conflicts arising from simultaneous execution are handled in the mapping module, so although each agent’s moves are processed independently, the code remains fully generalizable to any multi‐agent environment.

\begin{algorithm}[htbp]
\caption{Convert Grounded Trajectory to PDDL Trace Logs (\texttt{convert\_traj\_to\_pddl})}
\begin{algorithmic}[1]
\Require \texttt{trajectory}: list of timesteps, each containing a list of $(\text{agent},\text{action})$ pairs  
\Ensure $(\mathit{Grids},\mathit{Logs})$: sequence of grid states and action‐logs per timestep  
\State $\mathit{grid} \gets \text{InitGrid}()$  
\State $\mathit{Grids} \gets [\,]$; \quad $\mathit{Logs} \gets [\,]$  
\For{each timestep $t = 0$ \textbf{to} $|\text{trajectory}| - 1$}
    \State $\mathit{stepActions} \gets \text{trajectory}[t].\text{action}$
    \State $\mathit{logCurrent} \gets \{\}$
    \For{each $(agent, act) \in \mathit{stepActions}$}
        \State $(\mathit{pre}, \mathit{eff}, \mathit{del}, \mathit{grid}) \gets \text{ApplyAction}(act, \mathit{grid}, agent)$
        \State $\mathit{logCurrent}[agent] \gets \{\,
          \texttt{pre\_conditions}: \mathit{pre},\;
          \texttt{effects}:       \mathit{eff},\;
          \texttt{deletes}:       \mathit{del}\,\}$
    \EndFor
    \State Append $\bigl(\text{clone}(\mathit{grid})\bigr)$ to $\mathit{Grids}$
    \State Append $\mathit{logCurrent}$ to $\mathit{Logs}$
\EndFor
\State \Return $(\mathit{Grids},\mathit{Logs})$
\label{alg:convert_traj_to_pddl}
\end{algorithmic}
\end{algorithm}

\paragraph{Key Helper Functions:}
\begin{itemize}
  \item \texttt{ApplyAction(action, grid, agent\_index)}:  
    Applies the specified action for the given agent on the current grid state, returning  
    the pre-conditions, effects, deletes list, and the updated grid.  
    \textbf{Note:} This function is domain-dependent and must be implemented  
    according to the specific dynamics and action schema of your environment.
\end{itemize}
\subsection{Analysis Module}
The analysis module, as depicted in Algorithm 2, provides a domain-agnostic framework for detecting and categorizing interdependent interactions between agents within a multi-agent environment. Given a sequence of environment states (\texttt{snapshots}) and corresponding action logs parsed from PDDL traces (generated by the mapping module), the algorithm dynamically maintains effect lists for each agent. At each timestep, the algorithm systematically checks whether the preconditions of an agent's action are satisfied by the effects of another agent's prior actions, thereby identifying potential interdependencies. Each detected interdependence is further classified into constructive, looping, irrelevant, or non-constructive categories by evaluating whether the object involved contributes to a goal, is repeatedly exchanged, or is otherwise extraneous. This modular design enables the analysis code to be readily applied across different domains, provided that the environment logs have been mapped to a consistent PDDL schema by the~/ref[section:mapping]{mapping module}.
\begin{algorithm}[ht!]
\caption{Detecting Interdependencies and Their Types in the grounded state and action trajectory (\texttt{detect\_int})}
{\normalsize
\raggedright
\begin{algorithmic}[1]
\sloppy
\Require Data logs: \texttt{snapshots} (state log), \texttt{action\_logs}; 
\Ensure Counts of interdependencies along with their types, and lists of actions by each agent which triggered an interdependence.

\State \textbf{For each agent:} effect\_list[agent] $\gets$ [\,] \Comment{Initialize empty effect list}

\For{each timestep $t$ up to trajectory length}
    \For{each agent}
        \If{agent delivers an object}
            \State Record the delivered object in goal objects array
        \EndIf
    \EndFor
\EndFor

\For{each timestep $t$ up to trajectory length}
    \For{each agent}
        \State effect\_list[agent] $\gets$ \texttt{filter\_effect\_list\_by\_state}(effect\_list[agent], snapshots[t])
        \State Check if the current action's precondition matches an effect in the other agent's effect list via \texttt{check\_precondition\_in\_effect\_list}
        \If{precondition matches}
            \State Assess: \\
\State \parbox[t]{\dimexpr\linewidth-4em}{%
\textbf{Goal-reaching:} Is the object part of the goal? (\texttt{check\_if\_int\_goal})\\[0.5ex]
\textbf{Giver loop:} Does the object return to the giver in the same state? (\texttt{check\_if\_giver\_loop})\\[0.5ex]
\textbf{Receiver loop:} Did the receiver ever possess the object in the same state? (\texttt{check\_if\_receiver\_loop})%
}
            \If{all conditions met}
                \State Increment constructive interdependencies count
            \ElsIf{loops detected}
                \State Increment looping interdependencies count
            \ElsIf{not goal-reaching}
                \State Increment irrelevant interdependencies count
            \Else
                \State Increment non-constructive interdependencies count
            \EndIf
        \EndIf
    \EndFor
    \State Save deep copy of current effect lists for next timestep
\EndFor

\State \Return Interdependence counts of four types, list of trigger actions for each agent
\label{alg:detect_int}
\end{algorithmic}
}
\end{algorithm}
\paragraph{Key Helper Functions:}
\begin{itemize}
  \item \texttt{extract\_cells\_with\_object(grid\_state)}: Extracts cells containing an 'object' property.
  \item \texttt{filter\_effect\_list\_by\_state(effect\_list, state\_snapshot)}: Filters and deduplicates effect entries by verifying object presence and state against the snapshot.
  \item \texttt{check\_precondition\_in\_effect\_list(action, effect\_list\_other\_agent)}: Checks if an action's precondition matches any effect in another agent's effect list.
  \item \texttt{check\_if\_int\_goal(int\_obj\_id, goal\_object\_arr)}: Determines if an object is part of the goal.
  \item \texttt{check\_if\_giver\_loop(int\_obj\_id, giver\_agent\_id, snapshots)}: Checks if a giver receives the object back.
  \item \texttt{check\_if\_receiver\_loop(int\_obj\_id, rec\_agent\_id, snapshots)}: Checks if the receiver already held the object.
\end{itemize}
\label{section:analysis}
\section{Illustrating Evaluation of Cooperative Behavior in a Search and Rescue Domain}
\label{section:usar}
We demonstrate that the proposed metric for measuring cooperation generalizes naturally to a heterogeneous Search and Rescue (SAR) domain.
The domain simulates a common emergency setting—a house partially engulfed in flames with multiple victims scattered throughout. 
The scenario is modeled on a discrete 2D grid representing rooms and hallways within the house. Some areas are blocked by debris or actively burning fires, and victims may be located in proximity to these hazards. Successful rescue requires coordinated efforts from a heterogeneous team of agents — each with specialized capabilities and constraints.  With its heterogeneous team of firefighters and nurses, this domain provides a rich testbed for analyzing cooperative behavior.

\subsection{Domain Specification}

We define the SAR environment as:
\[
\mathcal{G}_{\text{SAR}} = \langle \mathcal{I}, \mathcal{S}, \mathcal{A}, T, R, \gamma \rangle
\]

\begin{itemize}[leftmargin=*]
  \item \textit{Agents:} \( \mathcal{I} = \{ \text{Nurse (N)}, \text{Firefighter (F)} \} \)
  \begin{itemize}
    \item \textbf{Nurse (N)}: Can treat victims without a medical kit as well as administer aid using a medical kit to victims.
    \item \textbf{Firefighter (F)}: Can extinguish fire using a fire extinguisher.
\end{itemize}
The locations and states of all the victims is unknown to the agents upon initialization. All the agents explore the space to discover new victims.
  \item \textit{State Space:} \( \mathcal{S} \) includes:
  \begin{itemize}
    \item \textbf{Agent Locations:} The grid coordinates of each agent.
    \item \textbf{Victim Locations:} The positions of all victims in need of rescue.
    \item \textbf{Victim Status:} Each victim may be in one of two states: \texttt{untreated} or \texttt{treated}.
    \item \textbf{Cell Conditions:} Each grid cell can contain:
    \begin{itemize}
        \item \texttt{Debris} (present or cleared),
        \item \texttt{Fire} (burning or extinguished).
    \end{itemize}
    \item \textbf{Agent Inventories:} For each agent, a list of carried objects (e.g., medical kit, fire extinguisher).
    \item \textbf{Guard Status:} A Boolean flag indicating whether an agent is currently being guarded by a police agent.
\end{itemize}
  \item \textit{Actions:} 
  Each agent has a discrete action space consisting of five actions: —up, down, left, and right and an interact action that allows it to engage with objects in the environment.
  
  \item \textit{Transition Function \(T\):} 
The environment transitions are governed by object-agent interactions and spatial constraints. The transition function $T(s, a, s')$ depends on the current state $s$, the agent’s action $a$, and environmental conditions. Some examples of critical transition functions in this domain are:
\begin{itemize}
    \item \textbf{Blocked Movement:} Movement actions are invalid or fail if the target cell contains uncleared debris or active fire.
    \item \textbf{ Interact(Firefighter, Extinguisher,Fire:} Fire in the target cell is extinguished.
    \item \textbf{Interact(Nurse,Medical Kit,Victim):} Victim status transitions from \texttt{untreated} to \texttt{treated} within 20 timesteps. It takes 100 timesteps if there is a fire in the room.
    \item \textbf{Interact(Firefighter,Medical Kit,Nurse:)} Transfers medical kit from firefighter to nurse.
    \item \textbf{Interact(Firefighter,Debris, Cell):} Clears debris in the current cell.
\end{itemize}

  \item \textit{Reward Function \(R\):} At the end of a run of a fixed number of timesteps, all agents receive +10 for each victim successfully treated.
  \end{itemize}
\subsection{Mapping to PDDL}
The Search and Rescue (SAR) domain described above can be seamlessly integrated with the \texttt{mapping} module to produce grounded symbolic trajectories. By specifying a domain configuration file, users can declaratively define the set of symbolic predicates (e.g., \texttt{VictimLocationKnown}, \texttt{Has(Nurse, MedicalKit)}, \texttt{FireExtinguished}), along with extraction functions that detect these predicates from environment states. Low-level actions, such as \texttt{Interact(Nurse, MedicalKit, Victim)}, are mapped to high-level symbolic operators with well-defined preconditions and effects. As agents traverse the environment and execute actions, the \texttt{mapping} module produces a symbolic trace that reflects the evolving state of the environment and the effects of agent actions, in a post-hoc manner. 
\begin{figure}[h]
    \centering
    \fbox{\includegraphics[width=1\linewidth]{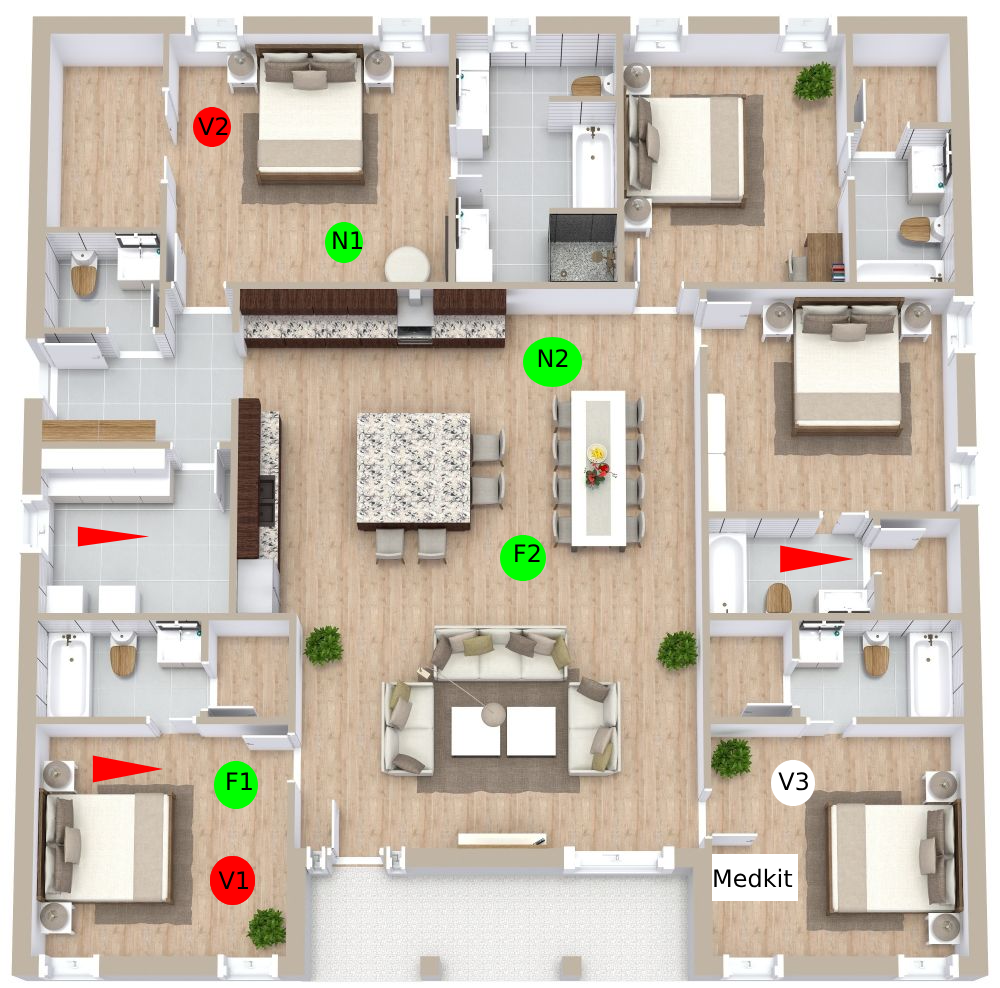}}
    \caption{Illustration of an instance of the Search and Rescue Domain}
\end{figure}

\subsection{Interdependencies in the SAR Domain}
Once trajectories are converted into grounded symbolic traces by the \texttt{mapping} module, the \texttt{analysis} module can be directly applied to detect and categorize interdependent interactions among agents. The analysis algorithm, as described in Algorithm~\ref{alg:detect_int}, processes these traces to dynamically track how agent actions influence one another. 
We can now formally define interdependencies between agents in the SAR domain. We illustrate examples of sequential interdependencies below:
\paragraph{Example 1: Firefighter discovers victim $\rightarrow$ Nurse treats victim : }
In this domain, firefighter and nurse agents collaboratively explore the environment to locate and assist victims. While they may search independently to maximize spatial coverage, coordination enables them to operate in parallel effectively. In this example, Firefighter~1 (F1) discovers Victim~1 (V1) by performing the action $a^{t_0}_j = \texttt{Interact(Firefighter, Victim)}$, which results in the predicate $\texttt{VictimLocationKnown} \in \text{add}(a^{t_0}_j)$. At the same time, Nurse~2 (N2) is exploring other areas. Once the victim's location is known, N2 can execute the action 

\[
a^{t_0 + k}_i = \texttt{Navigate}\left(
\begin{array}{l}
\texttt{Nurse Current Location,} \\
\texttt{Victim Location}
\end{array}
\right)
\]

which has $\texttt{VictimLocationKnown} \in \text{pre}(a^{t_0 + k}_i)$ as a precondition. Since only nurses are capable of treating victims, this coordination allows N2 to reach and assist V1. 

\begin{itemize}
     \item Giver Action:
\[
a^{t_0}_j = \texttt{Interact}\left(
\begin{array}{l}
\texttt{Firefighter,} \\
\texttt{Victim}
\end{array}
\right)
\]

    \item \text{Effect:} $\texttt{VictimLocationKnown} \in \text{add}(a^{t_0}_j)$
    \item Receiver Action:
\[
a^{t_0 + k}_i = \texttt{Navigate}\left(
\begin{array}{c}
\texttt{Nurse Current Location,} \\
\texttt{Victim Location}
\end{array}
\right)
\]

    \item \textbf{Precondition:} $\texttt{VictimLocationKnown} \in \text{pre}(a^{t_0 + k}_i)$
    \item \text{Object:} $\texttt{Victim}$
\end{itemize}
\paragraph{Example 2: Firefighter passes medical kit $\rightarrow$ Nurse treats victim : }
This scenario illustrates constructive sequential interdependence through the transfer of an object required for task completion. Nurse~1 (N1) needs a medical kit to treat Victim~2 (V2) but does not currently have one in their inventory and is located farther away from the kit. Firefighter~2 (F2), who is closer to the medical kit, performs the action $a^{t_0}_j = \texttt{Interact(Firefighter, MedicalKit, Nurse)}$, resulting in the effect $\texttt{Has(Nurse, MedicalKit)} \in \text{add}(a^{t_0}_j)$. This enables N1 to subsequently perform the action $a^{t_0 + k}_i = \texttt{Interact(Nurse, MedicalKit, Victim)}$, which has $\texttt{Has(Nurse, MedicalKit)} \in \text{pre}(a^{t_0 + k}_i)$ as a precondition. Since only nurses are capable of treating victims, F2’s assistance is critical in enabling N1 help V2. 

\begin{itemize}
    \item \text{Giver Action:}
    \[
    a^{t_0}_j = \texttt{Interact}\left(
    \begin{array}{l}
    \texttt{Firefighter,} \\
    \texttt{MedicalKit, Nurse}
    \end{array}
    \right)
    \]
    \item \text{Effect:} $\texttt{Has(Nurse, MedicalKit)} \in \text{add}(a^{t_0}_j)$
    \item \text{Receiver Action:}
    \[
    a^{t_0 + k}_i = \texttt{Interact}\left(
    \begin{array}{l}
    \texttt{Nurse,} \\
    \texttt{MedicalKit, Victim}
    \end{array}
    \right)
    \]
    \item \text{Precondition:} $\texttt{Has(Nurse, MedicalKit)} \in \text{pre}(a^{t_0 + k}_i)$
    \item \text{Object:} $\texttt{MedicalKit}$
\end{itemize}

\paragraph{Example 3: Firefighter extinguishes fire $\rightarrow$ Nurse treats victim faster : }
This example highlights constructive interdependence where one agent modifies the environment to improve the effectiveness of another agent’s action. In this scenario, Victim~3 (V3) is located in a room affected by fire, which hinders medical intervention. Nurse~2 (N2) is en route to treat the victim, but treatment is significantly faster and more effective if the fire has already been extinguished. Firefighter~1 (F1), who is in proximity to the fire, performs the action $a^{t_0}_j = \texttt{Interact(Firefighter, Extinguisher, Fire)}$, resulting in the effect $\texttt{FireExtinguished} \in \text{add}(a^{t_0}_j)$. This condition satisfies the precondition $\texttt{FireExtinguished} \in \text{pre}(a^{t_0 + k}_i)$ of the nurse’s treatment action $a^{t_0 + k}_i = \texttt{Interact(Nurse, MedicalKit, Victim)}$, thereby enabling faster and more efficient treatment. This form of interdependence ensures that F1’s timely intervention directly enhances N2’s ability to save the victim.

\begin{itemize}
    \item \text{Giver Action:}
    \[
    a^{t_0}_j = \texttt{Interact}\left(
    \begin{array}{l}
    \texttt{Firefighter,} \\
    \texttt{Extinguisher, Fire}
    \end{array}
    \right)
    \]
    \item \text{Effect:} $\texttt{FireExtinguished} \in \text{add}(a^{t_0}_j)$
    \item \text{Receiver Action:}
    \[
    a^{t_0 + k}_i = \texttt{Interact}\left(
    \begin{array}{l}
    \texttt{Nurse,} \\
    \texttt{MedicalKit, Victim}
    \end{array}
    \right)
    \]
    \item \text{Precondition:} $\texttt{FireExtinguished} \in \text{pre}(a^{t_0 + k}_i)$ (for fast treatment)
    \item \text{Object:} $\texttt{Fire}$
\end{itemize}

These sequential interdependencies are goal-reaching and non-looping.

\subsection{Looping vs. Non-Looping Sequential Interdependence}

In our framework, sequential interdependencies $(a_i^{t_0 + k}, a_j^{t_0})$ are defined as \textit{goal-reaching} if the interaction contributes to final reward acquisition (e.g., successful victim treatment), and \textit{non-looping} if the associated object $\text{obj}^{\text{int}}$ is not returned to the original agent in the same state. That is, the influence trajectory $\tau_{\text{obj}}^{t_0}$ must be strictly progressing toward a terminal effect and not cyclic with respect to the state of $\text{obj}^{\text{int}}$. To illustrate a \textit{looping interdependence}, consider the case where a police agent transfers a medical kit to a nurse at time $t_0$, and at time $t_0 + k$, the nurse returns the \textbf{same} kit to the police. If the state of the medical kit—denoted $s(\texttt{MedicalKit})$—remains unchanged (e.g., \texttt{unused}, \texttt{intact}, \texttt{full-capacity}), and the kit does not contribute to any further task completion, then this constitutes a \textit{looping} and \textit{non-goal-reaching} interdependence. It is redundant and does not affect the task reward.
Here, we consider a more nuanced scenario: At time $t_0$, the \textit{nurse agent} transfers a \texttt{MedicalKit} to the \textit{police agent} temporarily to free up their inventory (e.g., under an assumption that the nurse can initiate victim treatment bare-handed). At a later time $t_0 + k$, the police agent returns the same \texttt{MedicalKit} to the nurse, who then uses it to \textbf{complete} the victim treatment. In this case:
\begin{itemize}
    \item The interdependence is \textit{goal-reaching} since the treatment concludes successfully with enhanced reward.
    \item However, it is \textit{looping}, as the object returns to its original holder in the same nominal state.
\end{itemize}

To resolve the case of \textit{useful transfers}, we modify the state of the \texttt{MedicalKit} by augmenting it with a \textit{usage-linked attribute}, such as:
\[
s(\texttt{MedicalKit}) = 
\begin{cases}
\texttt{unused} \\
\texttt{used-for-treatment} \\
\texttt{passed-temporarily}
\end{cases}
\]
By tagging the medical kit’s state based on the context in which it was transferred (e.g., part of a treatment pipeline for a victim), we can distinguish \textit{constructive looping interdependence} from \textit{useful transfers}. This allows us to retain goal-relevant looping interdependencies while discarding non-contributing loops.

\section{User Study Design:}
We conducted a user study to evaluate the performance of state-of-the-art zero-shot coordination (ZSC) agents in a cooperative cooking game. The user study was built from ~\citet{cole, li2023tackling, sarkar2022pantheonrl}. The purpose of this study was to understand how well these AI agents coordinate with human partners in real-time gameplay. Below we describe the study design, participants, game environments, agent details, and data collection process.
\subsection{IRB Certification for this User Study}
\begin{figure}[h]
    \centering
    \fbox{\includegraphics[width=1\linewidth]{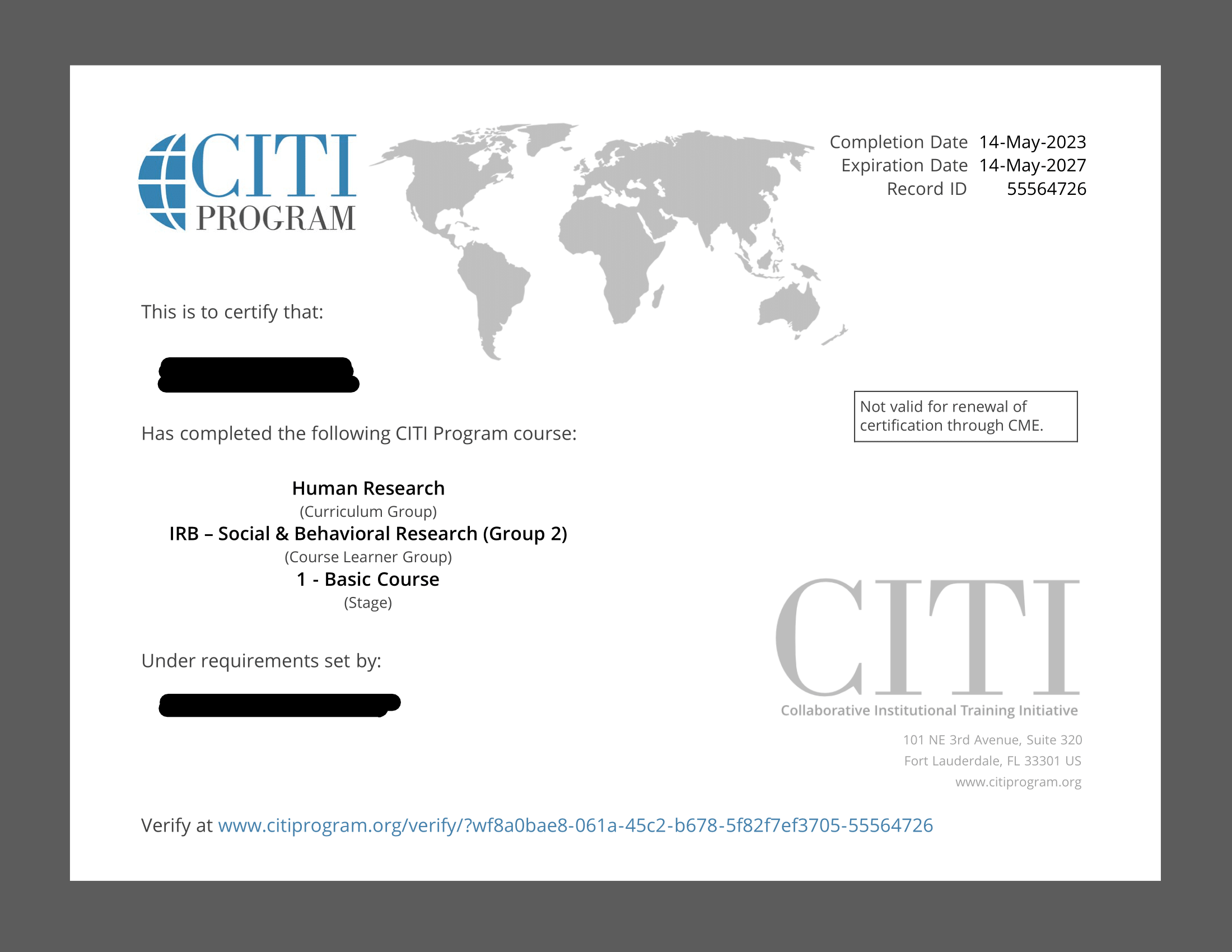}}
\end{figure}

\subsection{Consent and Experimental Statement}

Each participant began the study by reviewing and agreeing to a consent statement. The statement explained the goals of the study, what participants would be asked to do, and how their data would be handled.

\begin{itemize}
    \item \textbf{Purpose:} Participants were asked to take part in a study evaluating human performance when playing a cooperative cooking game with an AI partner.
    \item \textbf{Instruments:} The game was played using a computer screen and a keyboard.  
    \item \textbf{Procedure:}
    \begin{enumerate}
        \item After agreeing to the statement, participants filled out a demographic questionnaire.
        \item They read detailed instructions on how the game worked, including controls, rules, and objectives.
        \item They played a trial round with a scripted agent to become familiar with gameplay.
        \item They then played 16 rounds, each with a different pretrained AI partner.
        \item After each round, they filled out a short post-game questionnaire.
    \end{enumerate}
    \item \textbf{Confidentiality:} All data collected was kept confidential and anonymized. No personally identifiable information was stored or shared.
\end{itemize}

\begin{figure}[h]
    \centering
    \fbox{\includegraphics[width=1\linewidth]{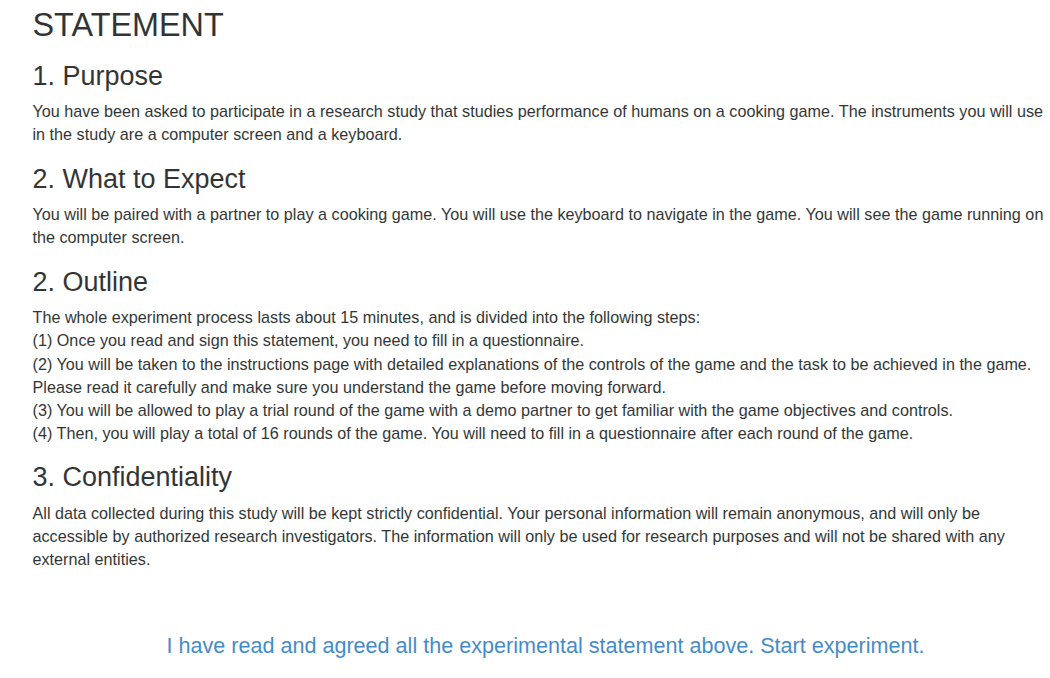}}
    \caption{Consent statement shown to participants before starting the study.}
\end{figure}

\subsection{Game Instructions and Layouts}

Participants were introduced to the game rules and controls through an instruction page. The game involves two players (one human, one AI), cooking and serving onion soup. Each round involved coordination to serve a single soup within 60 seconds.

\begin{figure}[h]
    \centering
    \fbox{\includegraphics[width=1\linewidth]{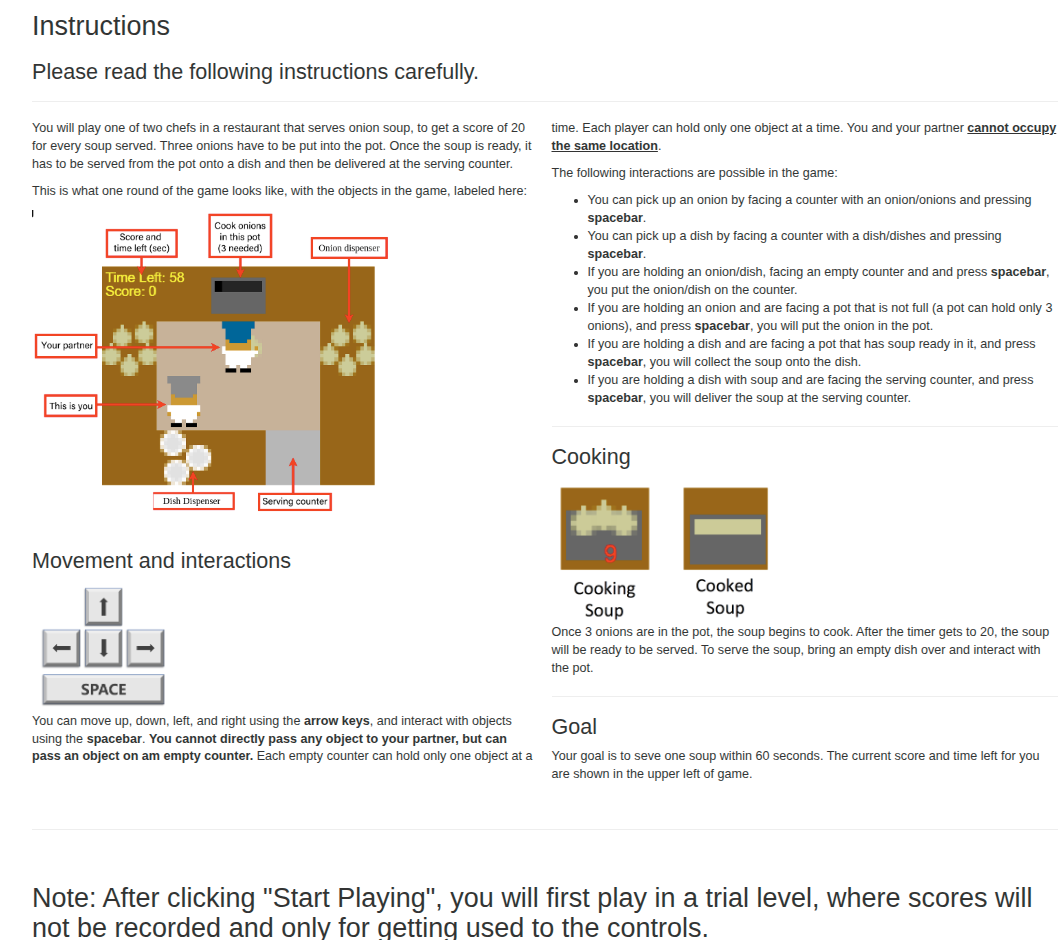}}
    \caption{Instructions page shown to participants.}
\end{figure}

We used two layout types in our evaluation:

\begin{itemize}
    \item Counter Circuit:  Players can perform independent tasks with minimal interference.
    \item Forced Coordination: A layout that restricts movement and requires players to coordinate, making collaboration essential.
\end{itemize}

\begin{figure}[h]
    \centering
    \fbox{\includegraphics[width=1\linewidth]{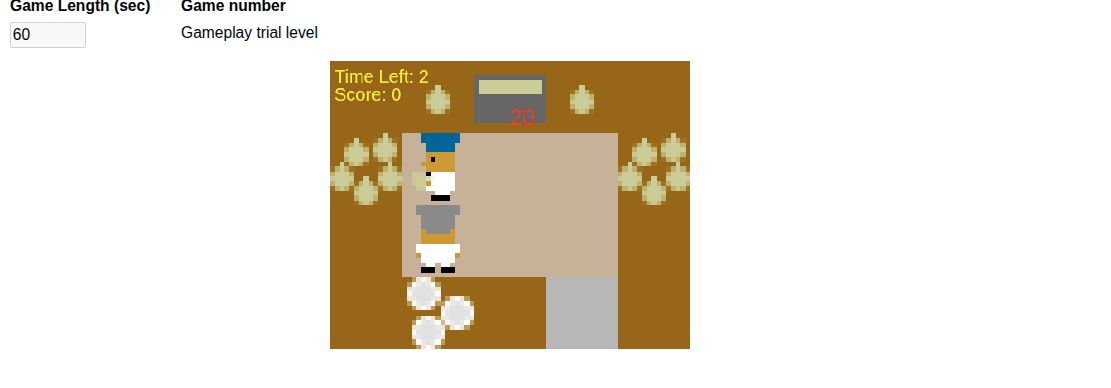}}
    \hfill
    \fbox{\includegraphics[width=1\linewidth]{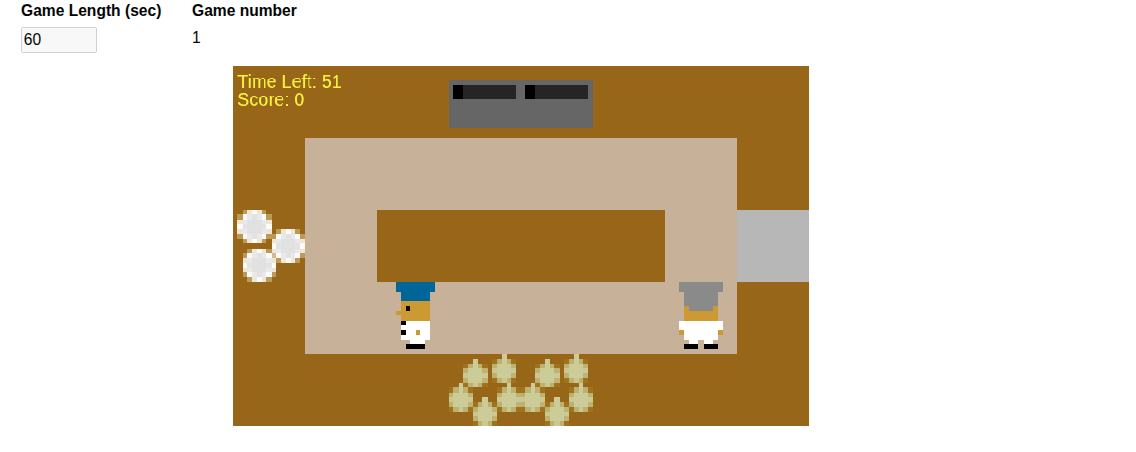}}
    \caption{Left: Trial round gameplay with scripted partner. Right: Real round gameplay with SOTA AI partner.}
\end{figure}

\subsection{AI Partners and Evaluation}

\textbf{SOTA Methods:} We evaluated four zero-shot coordination agents — FCP~\cite{fcp}, MEP~\cite{mep}, HSP~\cite{hsp}, and COLE~\cite{cole}. All these methods were trained using a two-stage framework:
\begin{itemize}
    \item \textbf{Stage 1:} A diverse partner population is created through self-play.
    \item \textbf{Stage 2:} The ego agent is trained by playing against sampled partners from the population and optimizing task rewards using reinforcement learning.
\end{itemize}

Each approach differs in how partner diversity is encouraged:
\begin{itemize}
    \item \textbf{FCP:} Direct self-play-based partner generation.
    \item \textbf{MEP:} Adds a maximum entropy term to encourage behavioral diversity in partners.
    \item \textbf{HSP:} Constructs agents that model human preferences using event-based rewards.
    \item \textbf{COLE:} Treats the game as a graphical-form cooperative game, with rewards based on cooperative incompatibility distributions.
\end{itemize}

The ego agent in each case is evaluated based on episodic task reward while paired with a human partner.

\subsection{Participants}

We recruited 36 participants aged between 18 to 31 from our university, with a median age of 22.5 and an average age of 20.75. Out of these, 24 participants identified as male and 12 as female. A majority (63.9\%) reported no prior familiarity with the Overcooked game. We conducted a pilot with 5 participants, and then used feedback from it to refine the final study with 31 participants.

\subsection{Post-Round Questionnaire}

After each round, participants filled out a questionnaire assessing collaboration, perceived responsiveness, and mutual intent. Each question was answered using a 5-point Likert scale (from "Strongly Disagree" to "Strongly Agree").

\begin{itemize}
    \item \textbf{Team Performance:}
    \begin{itemize}
        \item Q1. My partner and I worked together to deliver the soups.
        \item Q2. My partner contributed to the successful delivery of the soups.
    \end{itemize}
    \item \textbf{Were you working with your partner?}
    \begin{itemize}
        \item Q3. I attempted to work with my partner to deliver the soups.
        \item Q4. My partner responded to my attempts to work with them.
    \end{itemize}
    \item \textbf{Was your partner working with you?}
    \begin{itemize}
        \item Q5. My partner attempted to work with me.
        \item Q6. I responded to their attempts to work with them.
    \end{itemize}
\end{itemize}

\begin{figure}[h]
    \centering
    \fbox{\includegraphics[width=1\linewidth]{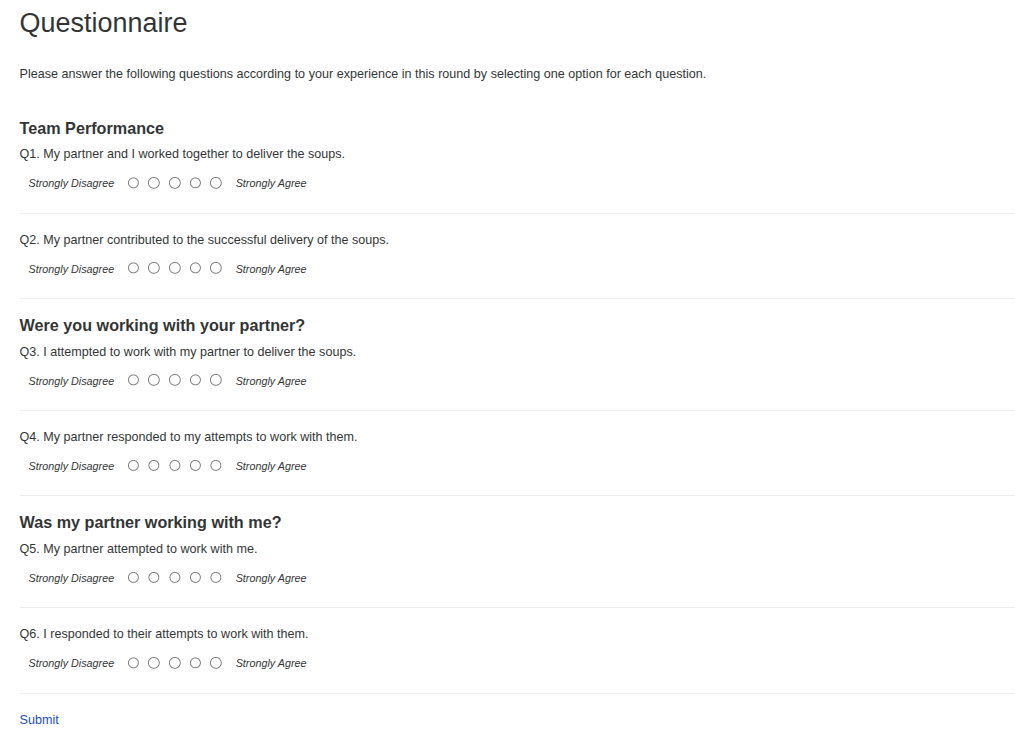}}
    \caption{Post-round questionnaire interface shown to participants.}
\end{figure}
\subsubsection{Questionnaire Design}
Evaluating teamwork purely through task-based performance (e.g., reward or completion time) can miss nuanced aspects of coordination, intent, and mutual understanding — particularly in zero-shot collaboration scenarios.
In this study, we developed an objective metric - \textbf{interdependence}, to measure the quality of team performance and cooperation in human-AI teams. 
The questionnaire was therefore designed to provide \textbf{additional subjective insights} into how humans perceived their AI partner's behavior. 
The central goal of our user study is to evaluate whether AI agents are capable of effective cooperation with human partners in zero-shot settings. Specifically, we want to assess two critical aspects of cooperative behavior:

\begin{itemize}
    \item \textbf{Responsiveness:} Does the AI agent recognize and respond to the human's attempts to collaborate?
    \item \textbf{Proactiveness:} Does the AI agent initiate behaviors that attempt to induce or enable cooperation from the human partner?
\end{itemize}
This enabled us to answer a core research question: \emph{Do the trajectories that score well under the interdependence metric also align with human perceptions of effective teamwork?} This alignment — or misalignment — between subjective and objective measures of teaming can reveal important gaps in AI-agent design, particularly in cooperative settings where behavior must be interpretable, responsive, and intuitive to humans.
Each questionnaire was administered after a single round of gameplay and asked participants to reflect on their experience with that round’s AI partner. The questions were grouped into three conceptual categories:

\begin{itemize}
    \item \textbf{Team Performance (Q1, Q2):} These items measure whether the participant felt the round involved joint effort and contribution from both teammates toward the goal of delivering soup.
    \item \textbf{Agent Responsiveness to Participant Coordination (Q3, Q4):}  Evaluates how effectively the agent responds when the participant initiates coordination.
    \item \textbf{Agent-Initiated Coordination and Participant Response (Q5, Q6):} Assesses how often the agent initiates coordination and how well these attempts are received by the participant.turn.
\end{itemize}

Each question was answered on a 5-point Likert scale (from \emph{Strongly Disagree} to \emph{Strongly Agree}). This design was inspired by constructs in human-robot interaction and team cognition research, such as perceived shared agency, responsiveness, and mutual intention. The repeated structure across 16 gameplay rounds allowed us to collect a rich set of human-AI interaction trajectories paired with subjective labels.
\subsubsection{Statistical Tests}

We tested the following null hypotheses related to participants' subjective perceptions of cooperation with their AI partners:

\begin{itemize}
    \item \textit{Counter Circuit Layout:}
    \begin{itemize}
        \item $H_0^{1.1}$: The mean response to the statement \textit{"My partner responded to my attempts to work with them"} equals the neutral midpoint (i.e., mean $= 3$).
        \item $H_0^{1.2}$: The mean response to the statement \textit{"My partner attempted to work with me"} equals the neutral midpoint (i.e., mean $= 3$).
    \end{itemize}
    \item \textit{Comparison Between Layouts (Counter Circuit vs. Forced Coordination):}
    \begin{itemize}
        \item $H_0^{2.1}$: There is no difference in mean responses to \textit{"My partner responded to my attempts to work with them"} between the two layouts (i.e., mean difference $= 0$).
        \item $H_0^{2.2}$: There is no difference in mean responses to \textit{"My partner attempted to work with me"} between the two layouts (i.e., mean difference $= 0$).
    \end{itemize}
\end{itemize}

Formally, these hypotheses were tested using one-sample $t$-tests against the neutral midpoint for individual layouts, and paired $t$-tests for within-subject comparisons across layouts.
For Q4, regarding partner responsiveness, responses in the Counter Circuit layout yielded a mean rating of 3.33. The one-sample $t$-test rejected the null hypothesis ($t(168) = 3.04$, $p = 0.0027$), indicating that participants perceived their partners as responding to their cooperation attempts at a level significantly above neutral. When comparing the two layouts within participants, a paired $t$-test showed a statistically significant difference ($t(23) = -2.24$, $p = 0.0352$), with higher perceived responsiveness reported in the Forced Coordination layout. Similarly, for Q5, which captures perceptions of partner initiative to cooperate, the Counter Circuit responses averaged 3.31. The one-sample $t$-test again rejected the null hypothesis of neutrality ($t(168) = 2.80$, $p = 0.0057$), suggesting that participants generally agreed their partners attempted to work with them. 

Taken together, these subjective ratings suggest that, on average, participants felt their AI partners both responded to and attempted to cooperate with them. However, it is important to contextualize these findings within the broader experimental setting of zero-shot cooperation, where agents were paired with human participants exhibiting diverse behaviors — some actively seeking cooperation, while others preferred to act independently. This is reflected in objective measures, such as the average value of $\%H^{\text{trig}}_{\text{tot-sub}}$, which reveal that not all human participants wanted to engage in the cooperative strategy. These results underscore a key limitation of relying solely on subjective reports to evaluate cooperation: although participants generally perceive that their partners respond and attempt to work with them, this perception does not necessarily indicate that human-agent teams actually follow cooperative strategies. Objective behavioral analyses demonstrate that these teams did not do cooperative strategies, highlighting the importance of complementing subjective feedback with rigorous quantitative metrics when assessing human-agent collaboration.

\end{document}